\documentclass[prd,aps,reprint,onecolumn,superscriptaddress,tightenlines,nofootinbib,eqsecnum,preprintnumbers,longbibliography,12pt]{revtex4-2}
\usepackage{bbm}
\usepackage{amsmath,amssymb}
\usepackage{physics}
\usepackage{dsfont}
\usepackage{graphicx}
\usepackage{hyperref}
\usepackage[dvipsnames]{xcolor}

\begin{document}


\title{Unavoidable Higgs coupling deviations in the $Z_2$-symmetric Georgi-Machacek model
}
\def\Carleton{Ottawa-Carleton Institute for Physics, Carleton University, Ottawa, ON K1S 5B6, Canada}

\author{Carlos Henrique de Lima}
\email{carloshenriquedelima@cmail.carleton.ca}
\author{Heather E.\ Logan}
\email{logan@physics.carleton.ca}
\affiliation{\Carleton}

\date{September 27, 2022}
\begin{abstract}
The $Z_{2}$-symmetric version of the Georgi-Machacek model does not possess a decoupling limit in which all the new particles can be made arbitrarily heavy, opening the possibility that the model can be \emph{entirely} excluded if experiments reveal no deviations from the Standard Model.  We explore this model, focusing on the part of parameter space in which the vacuum expectation value of the triplets, $\nu_\chi$, is small.  In the small-$\nu_\chi$ limit, the second custodial-singlet scalar field $S$ necessarily becomes very light and can contribute to the total width of the 125~GeV Higgs boson $h$ via $h \to SS$. We show that this process, together with LHC measurements of the $h \to \gamma\gamma$ rate, entirely excludes masses $m_S < m_h/2$ and thereby severely constrains the parameter space, setting an experimental lower bound $\nu_\chi \gtrsim 12.5$~GeV on the vacuum expectation value of the triplets.  This lower bound makes it impossible to avoid deviations from the Standard Model in the couplings of $h$ to fermion and vector boson pairs. We study the remaining parameter space after imposing constraints from direct searches for the additional Higgs bosons, and show that it is on the edge of being fully excluded at $95\%$ confidence level by LHC measurements of the 125~GeV Higgs boson's couplings.  Measurements of these couplings at the future high-luminosity run of the LHC will have sufficient precision to entirely exclude the model at $5\sigma$ if no deviations from the Standard Model are observed. 
\end{abstract}


\maketitle

\section{Introduction}
The discovery of the 125 GeV Higgs boson at the CERN Large Hadron Collider (LHC)~\cite{higgsdiscovery1,higgsdiscovery2} opened a new avenue to test the Standard Model (SM) through measurements of Higgs boson properties.  So far the results indicate that the SM remains in good agreement with experiments; while we know that the SM cannot be the full story, we also know that any corrections from new physics appear to be small.   Many different models have been proposed as extensions to the SM that take this into account. One of these extensions is the Georgi-Machacek (GM) model~\cite{GM1,GM2}, which extends the scalar sector of the SM by adding two SU(2)$_L$ triplet fields.

The GM model has several interesting features that motivate its proposal and its use as a benchmark in LHC searches.  One of them is the possibility to enhance the Higgs coupling to vector bosons compared to its strength in the SM~\cite{vector1,Falkowski:2012vh,vector2,vector3}.  This enhancement can appear at tree level only in scalar sectors containing representations of SU(2)$_L$ larger than the usual SM doublet.  Associated with this enhancement is the novel presence of doubly-charged, singly-charged, and neutral Higgs bosons $(H_5^{\pm\pm}, H_5^{\pm}, H_5^0)$ transforming as a fiveplet under the custodial symmetry, which couple to vector boson pairs with a strength proportional to the vacuum expectation values (vevs) of the SU(2)$_L$ triplet fields.  The general GM model has been extensively studied in the literature~\cite{litGM1,litGM2,litGM3,litGM4,ang4,ang5,ang7,unitarity,m3const}. 

In this paper, we focus on a constrained version of the GM model in which the scalar potential is invariant under a $Z_2$ symmetry under which the SU(2)$_L$-triplet scalar fields are odd.  Imposing the $Z_2$ symmetry eliminates two trilinear terms in the scalar potential.  This version of the GM model, which we will refer to as the Z2GM model, was in fact the original model written down by Chanowitz and Golden in Ref.~\cite{GM2}, in which they imposed the $Z_2$ symmetry for simplicity.  The Z2GM model was considered in all the pioneering phenomenological studies until the scalar potential trilinear terms were first written down by Aoki and Kanemura in Ref.~\cite{unitarity}.  One interesting feature of the Z2GM model is that the $Z_2$ symmetry eliminates the dimension-four operators by which the complex SU(2)$_L$-triplet scalar would generate a Majorana mass for the SM neutrinos, thereby eliminating the need to require that the Yukawa couplings of these operators be extremely tiny.   On the other hand, the spontaneous breaking of the $Z_{2}$ symmetry can be cosmologically dangerous since it can generate domain walls in the early universe.  This issue has been studied in Ref.~\cite{wall}; here we assume that, if domain walls are created in the early universe, they decay fast enough that they are not a problem.

The model also possesses a dark matter phase in which the SU(2)$_L$ triplets do not acquire vevs so that the lightest $Z_2$-odd state is stable.  In this paper, we will however consider only the phase in which the triplet vevs are nonzero and the $Z_2$ symmetry is spontaneously broken.\footnote{In our opinion the dark matter phase of the Z2GM model is not particularly well-motivated as a model for dark matter because the main motivation for the global symmetry structure of the GM model in the first place is the prevention of large contributions to the electroweak $\rho$ parameter.  These contributions are absent when the vevs of the SU(2)$_L$ triplets are zero, so there is no particular reason why a dark matter model involving SU(2)$_L$ triplets would require the restrictive global symmetry structure of the GM model.}

The most important feature of the Z2GM model for our purposes is that it does not possess a decoupling limit~\cite{litGM1}.  This stems from the fact that, in the absence of the trilinear terms forbidden by the $Z_2$ symmetry, the scalar potential involves only two dimensionful parameters, which can both be eliminated in favour of the vevs of the doublet and triplet scalar fields.  These vevs are in turn bounded from above by the $W$ and $Z$ boson mass relations.  The masses of all the physical scalars in the Z2GM model can then be expressed as proportional to these vevs multiplied by various combinations of scalar quartic couplings.  Imposing perturbative unitarity on the scalar quartic couplings then bounds the masses of all the additional Higgs bosons in the Z2GM model to lie below about 700~GeV~\cite{unitarity}.  This fact, together with the increasing precision in the Higgs coupling measurements, can potentially allow the model to be \emph{entirely} ruled out in future experiments.  A similar analysis~\cite{thdm1,thdm2} of the non-decoupling $Z_2$-symmetric version of the two-Higgs-doublet model has already entirely excluded that model.

A further curious feature of the Z2GM model, evident already in the mass formulas of Ref.~\cite{GM2} but first studied explicitly in Ref.~\cite{vector3}, is that in the limit that the SU(2)$_L$-triplet scalar vevs become very small, the mass of one of the custodial-singlet Higgs bosons (which we will call $S$) also becomes very small.  This opens the possibility of a new decay mode for the 125~GeV Higgs boson into a pair of these lighter states; i.e., $h \to SS$.  We show that this possibility is experimentally excluded by a combination of measurements of the $h$ total width and the $h \to \gamma\gamma$ decay rate (the latter indirectly constrains model parameters in a way that prevents the $hSS$ coupling from being dialed to zero).  This imposes a lower bound of $m_S > m_h/2$ and, due to the relationship between $m_S$ and the triplet vevs in the Z2GM model, also puts an \emph{experimental} lower bound on the triplet vevs, thereby guaranteeing nonzero deviations from the SM in the tree-level couplings of $h$ to fermion and vector boson pairs.  Based on projections of Higgs coupling measurement precision at the high-luminosity LHC (HL-LHC)~\cite{HLPROJATLAS}, we show that the surviving parameter space of the Z2GM model will be entirely excluded at $5\sigma$ if the measured central values of the Higgs couplings remain SM-like.

The phenomenological implications of experimental constraints on the Z2GM model, particularly from Higgs coupling measurements, were previously studied in Refs.~\cite{vector3,ang4,diphotonOTHER}.  Our analysis updates the pioneering 2012--13 studies of Refs.~\cite{vector3,ang4} with the latest available LHC Higgs coupling data.  It also extends the recent analysis in Ref.~\cite{diphotonOTHER} by including a detailed analysis of the $m_S < m_h$ mass hierarchy, which was not considered in Ref.~\cite{diphotonOTHER}.  Furthermore, we refine the analysis in Ref.~\cite{diphotonOTHER} by basing our Higgs coupling constraints on the model-independent ATLAS measurements of ratios of Higgs couplings~\cite{ATLAS2019,ATLAS2021}, rather than using fits to the couplings themselves, which were made by the LHC collaborations using model assumptions that do not hold in the Z2GM model.  As a cross-check we also analyze the constraints on the Z2GM model from the code HiggsSignals~\cite{HiggsSignals}, which implements the LHC Higgs signal strength measurements.

In addition to the Higgs coupling measurements, we also impose all available direct searches for additional neutral and charged Higgs bosons. We apply most of these searches by using the public code HiggsBounds~\cite{HiggsBounds} to exclude model points that violate the 95\% confidence level (CL) experimental exclusion in the most sensitive applicable search channel.  There are additional important search channels relevant to the Z2GM model which are not captured by HiggsBounds, in particular those involving the doubly-charged Higgs and involving Drell-Yan production of pairs of custodial-fiveplet states.  For these searches, we use the direct implementations within the public code GMCALC~\cite{gmcalc}. These additional channels are vector boson fusion (VBF) $H^{\pm \pm}_{5} \rightarrow W^{\pm} W^{\pm} \rightarrow \text{like-sign dileptons}$~\cite{H5WW1,H5WW2}, Drell-Yan $H^{\pm \pm}_{5} \rightarrow \text{like-sign dileptons}$~\cite{DYELLH51,DYELLH52,DYELLH53}, Drell-Yan $H_{5}^{0}H_{5}^{\pm}$ with $H_{5}^{0} \rightarrow \gamma \gamma$~\cite{DYELLH01,DYELLH02} and Drell-Yan $H_{5}^{++} H_{5}^{--} \rightarrow W^{+}W^{+}W^{-}W^{-}$~\cite{ATLASnewIM}.
We also include the indirect constraint from $b\rightarrow s \gamma$~\cite{m3const}, which excludes large values of the triplet vev; most of the parameter region thereby excluded is also excluded by the direct searches.

As we show, the combination of the direct search constraints with the Higgs coupling modifier ratios from ATLAS excludes the entire parameter space at 95\% CL; using instead the HiggsSignals fit for the Higgs couplings results in a small surviving region at this confidence level.  Both methods lead to a surviving parameter region at 99\% CL.  This surviving parameter region has significant Higgs coupling modifications which can be decisively tested at the HL-LHC and a light custodial singlet (below 200~GeV) which can be searched for at future lepton colliders~\cite{ILCREPO,scalarinlepton,scalarinleptonPUB}.
 
This paper is organized as follows. In Sec.~\ref{sec:model} we review the $Z_2$-symmetric version of the GM model. In Sec.~\ref{sec:singlet} we study the mass matrix for the two custodial singlets and elucidate the physics that drives one of them light when the triplet vev becomes very small. In Sec.~\ref{sec:phenoGMZ2} we examine the phenomenology of the 125~GeV Higgs boson decays to $SS$ and $\gamma\gamma$ and their dependence on the underlying model parameters. In Sec.~\ref{sec:expSTAT} we apply the experimental constraints, first for the region of parameter space in which $m_S < m_h/2$ in which we demonstrate that this region is entirely excluded, and then for the remaining parameter space.  We conclude in Sec.~\ref{sec:conc}.  Some details about the behaviour of the custodial singlet mass matrix and a comparison of the Z2GM model with the unconstrained GM model are given in Appendix~\ref{sec:appendix}.  Finally, the details of our implementation of the theoretical and experimental constraints on the Z2GM model and of the scans over parameter space are collected in Appendix~\ref{app:details}.

\section{$Z_2$-symmetric Georgi-Machacek model} \label{sec:model}

The Georgi-Machacek model is an extension of the scalar sector of the SM. Its scalar sector consists of the usual Higgs doublet ($\phi^{+}$, $\phi^{0}$) with hypercharge $Y=1$, a complex triplet ($\chi^{++}$, $\chi^{+}$, $\chi^{0}$) with hypercharge $Y=2$ and a real triplet ($\xi^{+}$, $\xi^{0}$, $\xi^{-}$) with hypercharge $Y=0$ (here $\xi^- = -\xi^{+*}$ and we use the convention $Q = T^3 + Y/2$). This model is minimal in the sense of not extending the gauge group of the SM and not using higher representations of SU(2)$_L$ beyond triplets~\cite{DYELLH53}.  The $\rho$ parameter is preserved by imposing a global SU(2)$_{L}\times $SU(2)$_{R}$ symmetry on the scalar sector of the model, which will break down to the custodial SU(2) symmetry upon electroweak symmetry breaking.\footnote{This global symmetry is explicitly broken by hypercharge interactions, and hence can be preserved only at tree level~\cite{Gunion:1990dt}.  This means that the model is better seen as an effective low-energy description of some UV model such as a composite Higgs scenario~\cite{GM1}.  Nevertheless, the loop-induced custodial symmetry breaking is quantitatively small enough that the tree-level custodial-symmetric GM model remains a useful effective theory for experimental purposes at the LHC~\cite{Blasi:2017xmc,rg2,Keeshan:2018ypw}.}

To make the global SU(2)$_{L}\times $SU(2)$_{R}$ symmetry manifest, we express the scalar fields as a bi-doublet $\Phi$ and a bi-triplet $X$,
\begin{align}
\Phi =  \begin{pmatrix}
\phi^{0 *} & \phi^{+} \\
-\phi^{+*} & \phi^{0}
\end{pmatrix} , \qquad 
X=  \begin{pmatrix}
\chi^{0 *} & \xi^{+} & \chi^{++} \\
-\chi^{+*} & \xi^{0} & \chi^{+} \\
\chi^{++*} & -\xi^{+*} & \chi^{0}
\end{pmatrix},
\end{align}
where $\xi^0$ is a real field and the rest are complex.  We also impose an additional $Z_2$ symmetry under which the bi-triplet is odd:
\begin{align}
\Phi \rightarrow \Phi \, , \qquad X \rightarrow -X \, .
\end{align} 
The $Z_2$ symmetry serves to eliminate the lepton-number-violating Yukawa couplings of the complex triplet field to the lepton doublets, which would give rise to a neutrino mass proportional to the vev of the complex triplet.  Since we will consider triplet vevs in excess of 10~GeV, this is desirable to avoid small neutrino Yukawa couplings of order $10^{-10}$.  The remaining fermion masses are generated through Yukawa couplings of the Higgs doublet as in the SM.

The most general renormalizable scalar potential obeying these global symmetries can then be written as\footnote{We use the parameterization of Ref.~\cite{litGM1}.  A translation table to the notations of other papers can be found in the appendix of that reference.}
\begin{align}\label{eq:pot} \nonumber
V(\Phi,X) &= \frac{\mu^{2}_{2}}{2}\Tr( \Phi^{\dagger}\Phi) + \frac{\mu^{2}_{3}}{2}\Tr( X^{\dagger}X)  + \lambda_{1} [\Tr( \Phi^{\dagger}\Phi)]^{2} +\lambda_{2}\Tr( \Phi^{\dagger}\Phi)\Tr( X^{\dagger}X) \\ 
&+\lambda_{3} \Tr(X^{\dagger}XX^{\dagger}X)  + \lambda_{4} [\Tr( X^{\dagger}X) ]^{2} - \lambda_{5} \Tr ( \Phi^{\dagger}\tau^{i}\Phi\tau^{j})\Tr(X^{\dagger} t^{i} X t^{j}).
\end{align}
Notice the absence of trilinear couplings that would involve an odd number of $X$ fields; these are forbidden in the Z2GM by imposing the $Z_2$ symmetry.  
The generators for the doublet are given in terms of the standard Pauli matrices, $\tau^{i}=\sigma^{i}/2$, and for the triplets they are the $3 \times 3$ representation,
\begin{align}
t^{1}= \frac{1}{\sqrt{2}}\begin{pmatrix}
0 & 1 & 0 \\
1 & 0 & 1 \\
0 & 1 & 0
\end{pmatrix}, \qquad 
t^{2} = \frac{1}{\sqrt{2}}\begin{pmatrix}
0 & -i & 0 \\
i & 0 & -i \\
0 & i & 0
\end{pmatrix}, 
\qquad t^{3} = \begin{pmatrix}
1 & 0 & 0 \\
0 & 0 & 0 \\
0 & 0 & -1
\end{pmatrix}.
\end{align}

The spontaneous symmetry breaking of the electroweak gauge group is achieved through the vevs of both the bi-doublet and the bi-triplet (the latter also spontaneously breaks the $Z_2$ global symmetry):
\begin{align}
\expval{\Phi} = \frac{\nu_{\phi}}{\sqrt{2}} \mathds{1}_{2\times 2}, \qquad \expval{X}= \nu_{\chi} \mathds{1}_{3\times 3} \, . 
\end{align}
The neutral components of the real and complex triplets must obtain the same vev to preserve the custodial symmetry at tree-level. Applying these vevs in the gauge sector yields the $W$ and $Z$ masses:
\begin{equation}
	M_W^2 = g^2 \nu^2/4, \qquad \qquad M_Z^2 = (g^2 + g^{\prime 2}) \nu^2 /4,
\end{equation}
where
\begin{align}
\nu^{2} = \frac{1}{\sqrt{2} G_F} &= \nu_{\phi}^{2} + 8 \nu_{\chi}^{2} \approx (246 \, \text{GeV})^{2}  \, .
\end{align}
The next step is to perform the field redefinition in the rest of the theory and decompose the neutral fields into real and imaginary components:
\begin{align} \nonumber
\phi^{0} &= \frac{\nu_{\phi}}{\sqrt{2}} + \frac{1}{\sqrt{2}} ( \phi^{0}_{R}+i\phi^{0}_{I}),  \qquad  \xi^{0} = \nu_{\chi} +\xi^0_R,   \qquad
\chi^{0} = \nu_{\chi} + \frac{1}{\sqrt{2}} ( \chi^{0}_{R} + i \chi^{0}_{I} ).
\end{align}

This change of variables is then applied to the potential. The minimization condition for this model is:
\begin{align} \label{eq:min1}
\pdv{V}{\nu_{\phi}} &= \nu_{\phi} \left[ \mu^{2}_{2} +4\lambda_{1}\nu_{\phi}^{2}+3(2\lambda_{2}-\lambda_{5})\nu_{\chi}^{2}   \right] = 0 \, , \\ \label{eq:min2}
\pdv{V}{\nu_{\chi}} &= \nu_{\chi} \left[ 3\mu^{2}_{3}+3(2\lambda_{2}-\lambda_{5})\nu_{\phi}^{2}+12(\lambda_{3}+3\lambda_{4})\nu_{\chi}^{2}  \right] =0 \, .
\end{align}
We can use Eqs.~\eqref{eq:min1} and \eqref{eq:min2} to write $\mu_{2}^2$ and $\mu_{3}^2$ in terms of the other couplings. This leaves us with 6 free parameters ($\lambda_{1},\lambda_{2},\lambda_{3},\lambda_{4},\lambda_{5},\nu_\chi$),\footnote{The triplet vev $\nu_{\chi}$ can alternatively be traded for the dimensionless mixing angle $s_H \equiv \sin\theta_H = 2 \sqrt{2} \nu_{\chi}/\nu$; we will later exhibit our results in both parameterizations.} with only $\nu_{\chi}$ being dimensionful, but having an upper bound $\nu_{\chi} \leq \nu/\sqrt{8}$ when all the contribution to $\nu$ comes from the triplets. We then trade $\lambda_{1}$ to enforce the 125~GeV Higgs to have its measured mass, meaning that we are left with 5 free parameters.

We next diagonalize the potential Eq.~\eqref{eq:pot} to the mass eigenbasis. We can expect, after the diagonalization, to have the following decomposition under the custodial $SU(2)$:
\begin{align}
(\textbf{2},\textbf{2}) = \textbf{1} \oplus \textbf{3}, \qquad (\textbf{3},\textbf{3})= \textbf{1} \oplus \textbf{3} \oplus \textbf{5}.
\end{align}
This means that, after the breaking, we expect to have 2 custodial singlets ($h,S$), 2 custodial triplets ($G,H_{3}$), and one custodial fiveplet ($H_{5}$). One of the custodial triplets ($G$) contains the Goldstone bosons which will be eaten by the gauge fields. The custodial triplet sector diagonalization is:
\begin{align} 
G^{+} &= c_{H} \phi^{+} + s_{H} \frac{1}{\sqrt{2}}(\chi^{+}+\xi^{+}) \, , \\
G^{0} &= c_{H} \phi^{0}_{I} +s_{H} \chi^{0}_{I} \, , \\
H_{3}^{+} &= -s_{H} \phi^{+} + c_{H} \frac{1}{\sqrt{2}} ( \chi^{+}+\xi^{+}) \, , \\
H_{3}^{0} &= - s_{H} \phi^{0}_{I} + c_{H} \chi^{0}_{I} \, ,
\end{align}
where $c_{H}$ and $s_{H}$ are: 
\begin{align}
c_{H} = \frac{\nu_{\phi}}{\nu} \, , \qquad \, s_{H} = 2\sqrt{2}\frac{\nu_{\chi}}{\nu} \, .
\end{align}
The custodial fiveplet diagonalization is:
\begin{align} 
H^{++}_{5} &= \chi^{++} \, , \\
H^{+}_{5} &= \frac{1}{\sqrt{2}} ( \chi^{+} - \xi^{+}) \, , \\
H^{0}_{5} &= \sqrt{\frac{2}{3}} \xi^{0}_R - \sqrt{\frac{1}{3}} \chi^{0}_{R} \, .
\end{align} 

The custodial symmetry at tree level enforces that the masses inside any custodial multiplet must be degenerate.  In the Z2GM, the masses of the custodial fiveplet and custodial triplet states are given respectively by:
\begin{align} \label{m5}
m_{5}^{2} &= \frac{3}{2} \lambda_{5} \nu_{\phi}^{2} + 8 \lambda_{3} \nu_{\chi}^{2} 
	= \left( \frac{3}{2} \lambda_5 c_H^2 + \lambda_3 s_H^2 \right) \nu^2 \, , \\
\label{m3} m_{3}^{2} &= \frac{1}{2} \lambda_{5} \nu^{2} \, .
\end{align}
Notice that the custodial triplet mass does not depend on $\nu_{\chi}$ (or equivalently, $s_H$), and is only sensitive to one quartic coupling.  The experimental lower bound on $m_3$ will thus set a lower bound on $\lambda_5$. The custodial fiveplet mass depends only on $\lambda_{3}$, $\lambda_{5}$ and $\nu_{\chi}$ and has the following relation in the $\nu_{\chi} \rightarrow 0$ (equivalently, $s_H \to 0$) limit:
\begin{align}\label{eq:rel}
m_{5} = \sqrt{3} m_{3} \qquad  \text{for} \, \, \nu_{\chi} \rightarrow 0 \, .
\end{align}

The remaining diagonalization is for the custodial singlet sector. Since this is the focus of the paper and the source of our most interesting results, we introduce the procedure here and develop it further in Section \ref{sec:singlet}. We first define the gauge basis as:
\begin{align}
H_{1}^{0} &= \phi^{0}_{R} \, , \\
H_{1}^{0'} &= \sqrt{\frac{1}{3}} \xi^{0}_R + \sqrt{\frac{2}{3}} \chi^{0}_{R} \, ,
\end{align}
where $H_1^{0'}$ is the custodial singlet that appears in the original bi-triplet.
These states mix to form the physical Higgs $h$, which we identify with the 125~GeV Higgs boson, and an additional CP-even custodial singlet $S$. The mass diagonalization can be done using the following orthogonal matrix:
\begin{align}\label{eq:rot}
\begin{pmatrix}
h \\
S 
\end{pmatrix} = \begin{pmatrix}
c_{\alpha} & -s_{\alpha}\\
s_{\alpha} & c_{\alpha}
\end{pmatrix} \begin{pmatrix}
H_{1}^{0} \\
H_{1}^{0'}
\end{pmatrix} \qquad ({\rm for} \ m_S > m_h).
\end{align}
Now we have to be careful with the definition of the angle $\alpha$. To be consistent with Refs.~\cite{litGM1,unitarity,vector2,vector3,ang4,ang5,ang7} and with GMCALC~\cite{gmcalc}, we define the rotation in terms of the heavy and light mass eigenstates. This means that for $m_{S} > m_{h}$ we have Eq.~\eqref{eq:rot}, while for $m_{S} < m_{h}$ we have:
\begin{align}\label{eq:rotlow}
\begin{pmatrix}
S \\
h 
\end{pmatrix} = \begin{pmatrix}
c_{\alpha} & -s_{\alpha}\\
s_{\alpha} & c_{\alpha}
\end{pmatrix} \begin{pmatrix}
H_{1}^{0} \\
H_{1}^{0'}
\end{pmatrix} \qquad ({\rm for} \ m_S < m_h).
\end{align}
Since we will want $h$ to be mostly $H_{1}^{0}$ for consistency with the LHC measurements of the Higgs couplings, this means that in the region where $m_S > m_h$ the SM limit is $c_{\alpha} \to 1$, while in the region where $m_{S} < m_{h}$ the SM limit is $s_{\alpha} \rightarrow 1$. 
In the next section, we analyze the custodial singlet sector further, highlighting, in particular, the region of parameter space in which one of the custodial singlets becomes very light.

In what follows we will perform numerical scans of the full parameter space of the Z2GM model, imposing the usual theoretical constraints (perturbative unitarity, boundedness from below, and absence of deeper custodial-violating minima) on the parameters of the scalar potential as implemented in GMCALC.  Details of these constraints, as well as of our scan procedure, are given in Appendix~\ref{app:details}.  We will also apply from the start the experimental lower bound on the custodial-triplet and -fiveplet masses~\cite{DYELLH53,LEPHiggsWorkingGroupforHiggsbosonsearches:2001ogs} (see Appendix~\ref{app:details} for details),
\begin{align}
	m_{3} , m_{5} \geq 76~\text{GeV} \, .
\end{align}
Perturbative unitarity of the quartic couplings in the scalar potential, together with the measured value of the Fermi constant $G_F$, also lead to upper bounds on these masses of $m_3 \lesssim 400$~GeV and $m_5 \lesssim 700$~GeV~\cite{unitarity}.

\section{Diagonalization of the custodial singlet sector} 
\label{sec:singlet}

\subsection{Features of the mass matrix}

In the custodial singlet sector, the mass matrix before diagonalizing has the form:
\begin{align} \label{eq:mass}
\mathcal{M}^{2} = \begin{pmatrix}
\mathcal{M}_{11}^{2} & \mathcal{M}_{12}^{2} \\
\mathcal{M}_{12}^{2} & \mathcal{M}_{22}^{2}
\end{pmatrix} \, ,
\end{align}
where:
\begin{align}
\mathcal{M}_{11}^{2} &= \mu_{2}^{2}+ 12 \lambda_{1} \nu_{\phi}^{2} + 3(2\lambda_{2}-\lambda_{5})\nu_{\chi}^{2} \, , \\
\mathcal{M}_{12}^{2} &= 2\sqrt{3} \nu_{\chi}\nu_{\phi} ( 2 \lambda_{2} - \lambda_{5}) \, , \\
\mathcal{M}_{22}^{2} &= \mu_{3}^{2}+12\nu_{\chi}^{2}(\lambda_{3}+3\lambda_{4} ) + (2\lambda_{2}-\lambda_{5})\, . \label{eq:M22before}
\end{align}
The minimization conditions in Eqs.~\eqref{eq:min1} and~\eqref{eq:min2} can be solved in the broken phase ($\nu_\phi, \nu_\chi \neq 0$) for $\mu_{3}^{2}$ and $\mu_{2}^{2}$, allowing the mass matrix to be expressed entirely in terms of vevs and quartic couplings:
\begin{align}\label{eq:M22beg}
\mathcal{M}_{11}^{2} &= 8 \lambda_{1} \nu_{\phi}^{2} \, , \\
\mathcal{M}_{12}^{2} &= 2\sqrt{3} \nu_{\chi}\nu_{\phi} ( 2 \lambda_{2} - \lambda_{5}) \, , \\
\mathcal{M}_{22}^{2} &= 8\nu_{\chi}^{2}(\lambda_{3}+3\lambda_{4} ) \, . \label{eq:M22}
\end{align}
It is clear from Eqs.~\eqref{eq:M22beg}--\eqref{eq:M22} that in the limit that $\nu_{\chi}$ approaches zero\footnote{We explicitly take this limit coming from positive nonzero values of $\nu_\chi$.  At $\nu_\chi = 0$ a phase transition occurs to the dark matter phase of the theory in which the $Z_2$ symmetry is unbroken; in this dark matter phase the SU(2)$_L$ triplets are $Z_2$-odd, their masses are controlled by $\mu_3^2$ as in Eq.\eqref{eq:M22before}, and they can be decoupled by taking $\mu_3^2 \gg \nu^2$.  We do not consider the dark matter phase in this paper.} the Higgs boson $h$ acquires a mass-squared of the SM form $8 \lambda_1 \nu^2$, the mixing goes to zero, and the second custodial singlet's mass approaches zero.  The small $\nu_{\chi}$ region is thus populated by a low-mass custodial singlet $S$. We defined the diagonalization matrix depending on the mass hierarchy in Eq.~\eqref{eq:rot} and  Eq.~\eqref{eq:rotlow}. For both cases the mixing angle $\alpha$ can be computed using:
\begin{align}
	\sin(2\alpha) &= \frac{2 \mathcal{M}^{2}_{12}}{|m_{h}^{2}-m_{S}^{2}|} \, , \\
	\cos(2\alpha) &= \frac{\mathcal{M}^{2}_{22}-\mathcal{M}^{2}_{11}}{|m_{h}^{2}-m_{S}^{2}|} \, .
\end{align}

Note that, after fixing $\nu$ using the Fermi constant, the custodial-singlet sector depends on only four combinations of parameters: $\nu_{\chi}$, $\lambda_{1}$, $(2\lambda_{2}-\lambda_{5})$ and $(\lambda_{3}+3\lambda_{4})$. We can simplify the analysis by defining the combinations:
\begin{align}
\lambda_{25} = 2\lambda_{2} - \lambda_{5}, \qquad \lambda_{34} = \lambda_{3} + 3 \lambda_{4} \, ,
\end{align}
so that
\begin{align}
	\mathcal{M}^2_{12} = 2 \sqrt{3} \nu_\chi \nu_\phi \lambda_{25}, \qquad
	\mathcal{M}^2_{22} = 8 \nu_\chi^2 \lambda_{34}.
\end{align}

Because the mass of $h$ is already measured to be 125~GeV, we use this to fix the value of $\lambda_{1}$. This inversion is independent of the mass hierarchy between $h$ and $S$:
\begin{align} \label{eq:l1}
\lambda_{1} = \frac{1}{8\nu_{\phi}^{2}} \left( m_{h}^{2} + \frac{(\mathcal{M}^{2}_{12})^{2}}{\mathcal{M}_{22}^{2} - m_{h}^{2}} \right) \, .
\end{align}
It is important to remember that $\lambda_{1}$ is constrained by perturbative unitarity\footnote{The upper bound on $\lambda_1$ from perturbative unitarity depends on the values of $\lambda_2$ and $\lambda_5$, but it cannot exceed $\pi/3$~\cite{litGM1}.  Boundedness-from-below of the scalar potential also constrains $\lambda_1 > 0$.} and this inversion may lead to a disallowed value for $\lambda_1$, in which case the point is discarded.

The variables $\lambda_{25}$ and $\lambda_{34}$ are constrained by perturbative unitarity and vacuum stability to lie in the ranges~\cite{litGM1}
\begin{align}
	\lambda_{25} \in \left( -\frac{8\pi}{3}, \frac{4\pi}{3} \right), \qquad
	\lambda_{34} \in \left( 0, \pi \right). 
\end{align} 
The lower bound on $\lambda_{34}$ can also be quickly obtained from the requirement that the physical masses-squared are positive: using the trace and determinant of the mass matrix,
\begin{align}\label{eq:trace}
{\rm Tr} (\mathcal{M}^{2}) &= m_{h}^{2} + m_{S}^{2} \, , \\
\label{eq:det} \det ( \mathcal{M}^{2}) &= m_{h}^{2} m_{S}^{2} \, ,
\end{align}
requiring that $m_S^2 > 0$ imposes that the determinant is positive, and as a consequence, $\lambda_{34}$ needs to be positive. 

The next step is to find an expression for $m_{S}$ in terms of these combinations of couplings that is independent of whether $m_S$ is greater or less than $m_h$. Using Eq.~\eqref{eq:l1} in Eq.~\eqref{eq:trace} we can find an expression for $m_{S}$:
\begin{align}\label{eq:ms}
m_{S}^{2} = 8\lambda_{34}\nu_{\chi}^{2} + \frac{16 \lambda_{25}^{2}\nu_{\phi}^{2}\nu_{\chi}^{2}}{8\lambda_{34}\nu_{\chi}^{2} - m_{h}^{2}} \, .
\end{align}
We can subtract the Higgs mass-squared from Eq.~\eqref{eq:ms} to find the hierarchy relation:
\begin{align} \label{eq:hierachy}
m_{S}^{2}-m_{h}^{2} = \mathcal{K} + \frac{16 \lambda_{25}^{2}\nu_{\chi}^{2}\nu_{\phi}^{2}}{\mathcal{K}} \, ,
\end{align}
where we define the combination of parameters:
\begin{align}
\mathcal{K}= 8\lambda_{34}\nu_{\chi}^{2} - m_{h}^{2} \, .
\end{align}

\begin{figure}[h!]
   \resizebox{0.6\linewidth}{!}{ \includegraphics{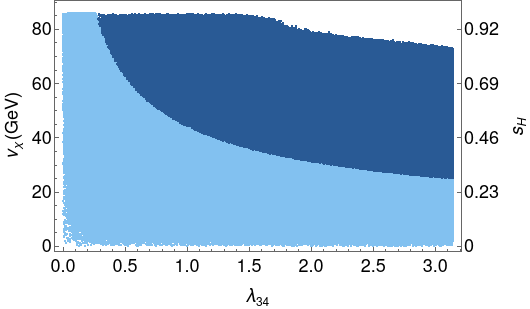}}
    \caption{\label{fig:l34byvx} The separation of the two possible mass hierarchies for the custodial singlets in the space of $\nu_\chi$ and $\lambda_{34}$, as controlled by the sign of $\mathcal{K}$. In dark blue we have points with $m_{S}>m_{h}$ ($\mathcal{K} > 0$) and in light blue we have points with $m_{S}<m_{h}$ ($\mathcal{K} < 0$). The boundary between these two regions is defined by $\mathcal{K}=0$. }
\end{figure}

We can see that the sign of $\mathcal{K}$ determines the hierarchy between the Higgs mass and the other custodial singlet mass. This happens because the numerator of the second term in Eq.~\eqref{eq:hierachy} is always positive. We generalize this result in Appendix~\ref{sec:appendix}, where we derive this relation for a general Hermitian matrix and also apply it to the general GM model with explicitly broken $Z_{2}$ to highlight the differences compared to the $Z_2$-symmetric version. 
When $\mathcal{K}$ is negative the Higgs $h$ is the heavier state, i.e., $m_S < m_h$. When $\mathcal{K}$ is positive the Higgs $h$ is the lighter state. We can see these two regions in Fig.~\ref{fig:l34byvx}. 

An additional relation can be found between $\lambda_{25}$ and $\lambda_{34}$ using the condition that the determinant is positive semi-definite (i.e., that $m_{S}^2 \geq 0$). Assuming that $\nu_{\chi} \neq 0$, we obtain a bound on how large $\abs{\lambda_{25}}$ can be:
\begin{align} \label{eq:satu}
16\lambda_{1}\lambda_{34} - 3 \lambda_{25}^{2} \geq 0 \, .
\end{align}
This imposes an additional upper bound on $\abs{\lambda_{25}}$. We can also express this condition using Eq.~\eqref{eq:ms} when $m_S < m_h$ (i.e., $\mathcal{K} < 0$) by requiring that $m_S^2 \geq 0$:
\begin{align}
\lambda_{25}^{2} \leq \frac{\lambda_{34}}{2\nu_{\phi}^{2}} |\mathcal{K}| 
= \frac{\lambda_{34}}{2 \nu_\phi^2} \left| 8 \lambda_{34} \nu_\chi^2 - m_h^2 \right| .
\end{align}
Choosing values for $\lambda_{34}$ and $\nu_{\chi}$, this puts an additional bound on $\lambda_{25}$. This relation is just the requirement that the lighter custodial-singlet mass-squared is positive semi-definite.

\subsection{The light custodial singlet region}

\begin{figure}[t!]
 \resizebox{0.6\linewidth}{!}{\includegraphics{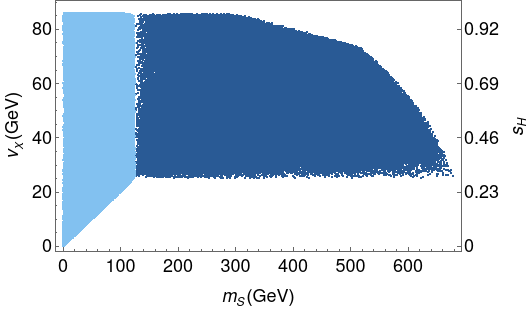}}
  \caption{\label{fig:massbyvxall} Allowed regions of $m_S$ versus $\nu_\chi$ after imposing the theoretical constraints and the mass bound $m_3, m_5 \geq 76$~GeV. We use the same color scheme as in Fig.~\ref{fig:l34byvx}, i.e., light blue points have $m_S < m_h$ and dark blue points have $m_S > m_h$. The region with $\nu_\chi$ below about 24.9~GeV is populated only by $m_S < m_h$. For values of $\nu_{\chi}$ smaller than about 12.5~GeV we have only points with $m_S < m_{h}/2$.}
\end{figure}

We now focus on the region with $m_S < m_h$.
In this region of the parameter space we have $\mathcal{K}$ negative, which from the definition of $\mathcal{K}$ means that:
\begin{align} \label{eq:light}
m_{h}^{2} > 8\lambda_{34} \nu_{\chi}^{2} \, .
\end{align}
If we make $\lambda_{34}$ in Eq.~\eqref{eq:light} as large as possible subject to perturbative unitarity and bounded-from-below constraints (i.e., $\lambda_{34}^{\rm max} = \pi$), we obtain that for $\nu_{\chi} < m_h/\sqrt{8 \pi} \simeq 24.9~\text{GeV}$ we \emph{always} have $m_S < m_h$. This value of $\nu_\chi$ is controlled by $\lambda_{34}$ alone: if the upper bound on $\lambda_{34}$ is reduced, then the value of $\nu_\chi$ below which only the hierarchy $m_S < m_h$ is possible becomes larger. We can see this interesting region in Fig.~\ref{fig:massbyvxall} in which we show the entire parameter space as a function of $\nu_\chi$ and $m_S$.  Again, the light blue points have $m_S < m_h$ and the dark blue points have $m_S > m_h$.  There is no overlap of the two colours in Fig.~\ref{fig:massbyvxall}.  From this figure, we see clearly that a lower bound on $m_S$ will translate directly into a lower bound on $\nu_\chi$ according to $\nu_\chi^{\rm min} = m_S^{\rm min}/\sqrt{8\pi}$ as long as $m_S^{\rm min} \leq m_h$.

At this point, we can understand why the authors of Ref.~\cite{diphotonOTHER} obtained a lower bound on $\nu_{\chi}$ by requiring that $m_S > m_h$. The low $\nu_{\chi}$ region is \emph{only} populated by the mass hierarchy $m_S < m_h$. This can be seen intuitively if we look at the custodial-singlet mass matrix in Eq.~\eqref{eq:mass}. Low values of $\nu_{\chi}$ mean that the off-diagonal terms and the $(2,2)$ element are small. These elements of the matrix determine the mass of $S$, which thus becomes small as well. If we take $\lambda_{25}=0$, so that the off-diagonal terms vanish, this is even more evident since then $\mathcal{M}_{22}^{2} = 8 \nu_\chi^2 \lambda_{34}$ is identified with $m_S^2$ and it is proportional to $\nu_{\chi}^{2}$. 

An \emph{experimental lower bound} on $\nu_{\chi}$ can therefore be achieved if we can set a lower bound on $m_S$. For the phenomenology of this model, a lower bound on $\nu_{\chi}$ has important implications for all the couplings of the theory, in particular those of the 125~GeV Higgs boson $h$.  

In what follows we will demonstrate that the entirety of the parameter space with $m_S < m_h/2$ is excluded by a combination of the experimental constraints on $h \to SS$ and $h \to \gamma\gamma$, thereby setting a lower bound $\nu_\chi \gtrsim 12.5$~GeV, or equivalently $s_H \gtrsim 0.143$.  We first examine how the underlying parameters control these two Higgs decay observables and how the limited parameter freedom makes it impossible to accommodate them both simultaneously.

\section{Higgs decay phenomenology in the light singlet region}
\label{sec:phenoGMZ2}

In this section, we examine the underlying model parameters that control the 125~GeV Higgs boson $h$ decays to $SS$ in the region $m_S < m_h/2$ and the contributions of the singly- and doubly-charged Higgs bosons to the loop-induced $h \to \gamma\gamma$ decay.  Focusing on the small-$\nu_\chi$ region, we show that there is not enough parameter freedom in the Z2GM model to simultaneously tune the $h \to SS$ width to be sufficiently small and the $h \to \gamma\gamma$ width to be sufficiently SM-like to be able to satisfy the experimental constraints.

\subsection{Higgs decay to $SS$} 
\label{subsec:hdecay}

Assuming that the second custodial singlet is lighter than $m_{h}/2$, it is possible to have an additional decay channel for the 125~GeV Higgs boson, $h \rightarrow SS$. The Lagrangian that controls this process can be written as:
\begin{align}
\mathcal{L}_{hSS} = - \frac{g_{hSS}}{2} h S^{2} \, ,
\end{align}
where the relevant coupling (for $m_S < m_h$) is:
\begin{align}
g_{hSS} = 2s_{\alpha}\nu_{\phi} \left[ 12 c_{\alpha}^{2}\lambda_{1} + \lambda_{25}(s_{\alpha}^{2}-2c_{\alpha}^{2})\right] + 2\sqrt{3}c_{\alpha} \nu_{\chi} \left[\lambda_{25}(c_{\alpha}^{2}-2s_{\alpha}^{2}) +4s_{\alpha}^{2}\lambda_{34}  \right]       \, .
\end{align}
Here $\lambda_{25}$ and $\lambda_{34}$ are the same combinations of couplings that appear above in the custodial singlet sector mass matrix.

Since we will be primarily interested in the low $\nu_{\chi}$ region, we can understand the behaviour of the model in this region analytically by examining the residual coupling in the $\nu_{\chi} \rightarrow 0$ limit:\footnote{Remember that from the diagonalization matrix the SM limit in the $m_S < m_h$ region is $s_{\alpha} \rightarrow 1$.}
\begin{align}
	g_{hSS} \to 2\lambda_{25}\nu \qquad (\rm{for} \ \nu_\chi \to 0).
\end{align}
Because $\nu \simeq 246$ GeV, the coupling $\lambda_{25}$ would need to be extremely small to satisfy the experimental bounds on the $h$ total width when $m_S < m_h/2$. We can write the decay rate for this channel as
\begin{align}
\Gamma_{hSS} =|g_{hSS}|^{2} \frac{1}{32\pi m_{h}}\left( 1 - \frac{4m_{S}^{2}}{m_{h}^{2}}\right)^{1/2} \, ,
\end{align}
and add it to the total width of the Higgs into SM channels, where we take the SM value to be $\Gamma_{SM} = 4.09$~MeV~\cite{LHCHiggsCrossSectionWorkingGroup:2016ypw}:  
\begin{align} \label{kapo}
\Gamma_{T} &= \kappa_{H}^{2}\Gamma_{SM} + \Gamma_{hSS} \, .
\end{align}
In Eq.~\eqref{kapo}, $\kappa_{H}^{2}$ encodes the modifications to the decay widths to SM final states. We can use $\kappa_{H} \to 1$\footnote{The tree-level couplings of $h$ to fermion and vector boson pairs, and its loop-induced coupling to gluon pairs, indeed go to their SM values in the $\nu_\chi \to 0$ limit.  The loop-induced coupling of $h$ to photon pairs and to $Z\gamma$ do not approach their SM values in this limit due to the presence of additional light singly- and doubly-charged scalars; however, the contributions of these decays to the total Higgs width are so small (less than a percent) that we ignore them in this qualitative analysis.  We do of course include them in our full numerical analysis.} for $\nu_{\chi} \rightarrow 0$ to understand the expected behaviour in this limit. 

A generic constraint can be obtained from the indirect bound on the Higgs total width from an analysis of on- and off-shell $ZZ$ production~\cite{CMS2019} (see Appendix~\ref{app:details} for details).  Taking a conservative 99.7\% confidence level ($3\sigma$) constraint, we have $\Gamma_{T} < 19.1~\text{MeV}$. 
Setting $\kappa_{H} \rightarrow 1$ in the $\nu_\chi \to 0$ limit we thus get an upper bound on the allowed decay width for the $h \to SS$ channel of $\Gamma_{hSS} < 15.0~\text{MeV}$.\footnote{This is an extremely conservative constraint since such a large new contribution to the Higgs total width would suppress the Higgs branching ratios to all SM final states by more than a factor of 4, resulting in strong disagreement with experimental data.}
This translates into an upper bound at $99.7\%$ CL for $\abs{\lambda_{25}}$ in the $\nu_\chi \to 0$ limit of:
\begin{align}
	\abs{\lambda_{25}} \leq 0.028 \, .
\end{align}
Moving away from $\nu_{\chi} \rightarrow 0$ we will see that $\lambda_{25}$ still needs to be relatively small. 

\begin{figure}[t!]
 \resizebox{0.60\linewidth}{!}{\includegraphics{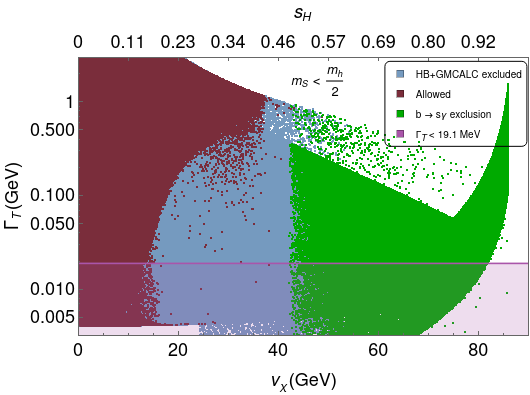}}
  \caption{\label{fig:higgsdecayconst} Total width of the 125~GeV Higgs as a function of the triplet vev $\nu_{\chi}$ from a scan over the Z2GM model parameter space in which $m_S < m_h/2$. The green points are excluded by experimental constraints on $b \to s \gamma$.  The light blue points are excluded by direct searches for the non-SM Higgs bosons as encoded in HiggsBounds and GMCALC (see Appendix~\ref{app:details} for details).  The maroon points are allowed after considering these constraints. The pink band at the bottom of the plot indicates the allowed range $\Gamma_T < 19.1$~MeV at 99.7\% CL.}
\end{figure}

In Fig.~\ref{fig:higgsdecayconst} we show the total width of $h$, scanning over the entire Z2GM parameter space for which $m_S < m_h/2$.  We can see that generic parameter values typically lead to enormous $h \to SS$ decay widths, but that a well-populated region of parameter space nevertheless exists in which the $hSS$ coupling is sufficiently suppressed to satisfy the indirect constraint $\Gamma_T < 19.1$~MeV.

\subsection{Higgs to diphoton decay}
\label{subsec:hdiphoton}

In the Z2GM model, the 125~GeV Higgs to diphoton decay rate is modified by the presence of the singly- and doubly-charged scalars $H_3^{\pm}$, $H_5^{\pm}$ and $H_5^{\pm\pm}$ running in the loop, as well as by modifications of the couplings of the Higgs to $W$ boson and fermion pairs. We adopt the usual normalization,~\cite{Gunion:1989we}
\begin{align}
\mathcal{L}_{int} =  -\frac{g m_{f_{i}}}{2M_{W}} \kappa_{f_{i}} \bar{f}_{i} f_{i}h + g M_{W} \kappa_{W} W^{+}W^{-}h - \frac{g m_{H_{i}}^{2}}{M_{W}}\kappa_{H_i} H_{i} H_{i}^{*} h \, ,
\end{align}
where the $\kappa_i$ parameterize the couplings of each particle to the 125~GeV Higgs boson (in the SM, $\kappa_W = \kappa_f = 1$).
This parameterization yields the expression for the partial width of $h \to \gamma\gamma$,~\cite{Gunion:1989we}
\begin{align}
	\Gamma(h \rightarrow \gamma \gamma) = \frac{\alpha^{2} g^{2} m_{h}^{3}}{1024\pi^{3}M_{W}^{2}} \abs{ \sum_{i} N_{i} Q_{i}^{2} \kappa_{i} F_{i}(\tau_{i})}^2 \, ,
\end{align} 
where $N_{i}$ is the number of colours of particle $i$, $Q_i$ is its electric charge in units of $e$, and $F_{i}(\tau_{i})$ is a loop function that depends on the spin of particle $i$ according to:~\cite{Gunion:1989we}
\begin{align}
F_{1}(\tau_{i})& = 2 + 3\tau_{i} + 3\tau_{i} \left( 2-\tau_{i} \right)f(\tau_{i}) \, , \\
F_{1/2}(\tau_{i}) &= -2\tau_{i} \left[ 1+(1-\tau_{i})f(\tau_{i})\right] \, , \\
F_{0}(\tau_{i}) &= \tau_{i} \left[ 1-\tau_{i}f(\tau_{i})\right] \, , \\
\end{align}
with $\tau_{i} = 4 m_{i}^{2}/m_{h}^{2}$ and
\begin{align}
f(\tau_{i}) &= \begin{cases} 
      \left[\sin^{-1}\left(\sqrt{\frac{1}{\tau_{i}}}\right)  \right]^{2} & \ {\rm for} \ \tau_{i} \geq 1, \\
      -\frac{1}{4}\left[\ln(\frac{1+\sqrt{1-\tau_{i}}}{1-\sqrt{1-\tau_{i}}}) -i\pi\right]^{2} & \ {\rm for} \ \tau_{i} < 1. 
   \end{cases}
\end{align}
 
We can then define an effective coupling modification factor $\kappa_\gamma$ for this decay amplitude relative to the SM:
\begin{align}
\kappa_{\gamma} = \frac{ \sum_{GM} N_{i}Q_{i}^{2}\kappa_{i}F_{i}(\tau_{i})}{ \sum_{SM} N_{i}Q_{i}^{2}F_{i}(\tau_{i})} \, ,
\end{align}
such that $\kappa_{\gamma}^2 = \Gamma_{GM}(h \to \gamma\gamma)/\Gamma_{SM}(h \to \gamma\gamma)$.
Writing out the couplings explicitly for the Z2GM model yields the expression:
\begin{align}
\kappa_{\gamma} = \frac{\kappa_{V}F_{1}(\tau_{W}) + \frac{4}{3}\kappa_{f}F_{1/2}(\tau_{t}) + \kappa_{3}F_{0}(\tau_{3}) + 5 \kappa_{5} F_{0}(\tau_{5})    }{F_{1}(\tau_{W}) + \frac{4}{3}F_{1/2}(\tau_{t}) } \, ,
\end{align}
where in both the Z2GM and SM amplitudes we have kept only the dominant fermionic contribution coming from the top quark.  We have also used the fact that $\kappa_W = \kappa_Z \equiv \kappa_V$ in the Z2GM model and that the $H_5^+ H_5^- h$ and $H_5^{++} H_5^{--} h$ couplings are controlled by the same coupling factor $\kappa_5$.

The expressions for the $\kappa$'s can be read directly from the Lagrangian by matching the definitions:
\begin{align}
g_{hW^{+}W^{-}} &= g M_{W} \kappa_{W} \, , \qquad g_{hff} = g \frac{m_{f}}{2M_{W}} \kappa_{f} \, , \\
g_{hH^{+}_{3}H^{-}_{3}} &= g \frac{m_{3}^{2}}{M_{W}} \kappa_{3} \, , \qquad 
g_{hH^{+}_{5}H^{-}_{5}} = g \frac{m_{5}^{2}}{M_{W}} \kappa_{5} \, , \qquad 
g_{hH^{++}_{5}H^{--}_{5}} = g \frac{m_{5}^{2}}{M_{W}} \kappa_{5} \, .
\end{align}

Because of the definitions of the mixing angle $\alpha$, the explicit expressions for the $\kappa$ factors depend on the mass hierarchy between $S$ and $h$.  Since in this section we are interested in the light singlet region, we give here the formulas valid for $m_S < m_h$.  The case of $m_S > m_h$ can be obtained from these by the replacements $s_\alpha \to c_\alpha$, $c_\alpha \to -s_\alpha$.  The gauge and fermionic couplings are:
\begin{align}
\kappa_{V} = s_{\alpha} \frac{\nu_{\phi}}{\nu} + \frac{8}{\sqrt{3}} c_{\alpha} \frac{\nu_{\chi}}{\nu} \, , \qquad \kappa_{f} = s_{\alpha}\frac{\nu}{\nu_{\phi}}. \qquad \qquad ({\rm for} \ m_S < m_h)
\end{align}
The (dimensionless) coupling factors of $h$ to charged custodial-triplet and -fiveplet scalar pairs are:
\begin{align}  \nonumber
\kappa_{3} &= \frac{1}{6m_{3}^{2}\nu} \Bigg[ 3s_{\alpha} \nu_{\phi}^{3}(4\lambda_{2}-\lambda_{5}) 
	+ 8\sqrt{3} c_{\alpha} \nu_{\phi}^{2}\nu_{\chi}(\lambda_{3}+3\lambda_{4}+\lambda_{5})  \\ 
\label{eq:kap3}
& \qquad \qquad \quad + 24s_{\alpha} \nu_{\phi}\nu_{\chi}^{2}(8\lambda_{1}+\lambda_{5}) 
	+ 16\sqrt{3}c_{\alpha} \nu_{\chi}^{3}(6\lambda_{2}+\lambda_{5})       \Bigg] \, , \\ \label{eq:kap5}
\kappa_{5} &= \frac{\nu}{2m_{5}^{2}} \left[  s_{\alpha} \nu_{\phi} (4 \lambda_{2} + \lambda_5) 
	+ 8\sqrt{3} c_{\alpha} \nu_{\chi}(\lambda_{3}+\lambda_{4}) \right]. \qquad \qquad 
		({\rm for} \ m_S < m_h).
\end{align}
The expressions for $\kappa_3$ and $\kappa_5$ depend on the masses $m_3$ and $m_5$, which are themselves dependent upon the underlying quartic couplings and vevs as given in Eqs.~\eqref{m3} and \eqref{m5}, respectively.  To connect the analysis of this sector to the preceding calculations, we define two additional linear combinations of quartic scalar couplings,
\begin{align}
\bar{\lambda}_{25} = 4\lambda_{2} + \lambda_{5}, \qquad 
\bar{\lambda}_{34}= \lambda_{3} + \lambda_{4}.
\end{align}
These are linearly independent from the combinations $\lambda_{25} = 2\lambda_{2} - \lambda_{5}$ and $\lambda_{34} = \lambda_{3}+3\lambda_{4}$ that we defined before.  Using these new couplings we can write the masses $m_3$ and $m_5$ as:
\begin{align} \label{eq:masses}
m_{3}^{2} &= (\bar{\lambda}_{25} -2\lambda_{25}) \frac{\nu^{2}}{6} \, , \qquad  m_{5}^{2} = (\bar{\lambda}_{25} -2\lambda_{25})\frac{\nu_{\phi}^{2}}{2} + 4( 3\bar{\lambda}_{34} -\lambda_{34})\nu_{\chi}^{2} \, .
\end{align}

We can then study the limit $\nu_{\chi} \rightarrow 0$ (which also implies $s_{\alpha} \to 1$) to gain some intuition about the behaviour of $\kappa_{\gamma}$. In this limit, $\kappa_V$ and $\kappa_f$ both go to 1, i.e., the tree-level gauge and Yukawa couplings of $h$ become SM-like. The deviation of $\kappa_{\gamma}$ from unity is then dependent solely on the contributions of the singly- and doubly-charged scalars in the loop:
\begin{align}
\kappa_{\gamma} \to 1 +  \frac{\kappa_{3}F_{0}(\tau_{3}) + 5 \kappa_{5} F_{0}(3\tau_{3})}{F_{1}(\tau_{W}) + \frac{4}{3} F_{1/2}(\tau_{t})} \, ,  \qquad ({\rm for} \ \nu_\chi \to 0)
\end{align}
where we also used the fact that in this limit $m_5 \to \sqrt{3} m_3$ as in Eq.~\eqref{eq:rel}. Applying this limit to Eqs.~\eqref{eq:kap3} and \eqref{eq:kap5} we get:
\begin{align}
\kappa_{3} \to \frac{\bar{\lambda}_{25} + 4 \lambda_{25}}{\bar{\lambda}_{25} - 2 \lambda_{25}} \, , \qquad
\kappa_{5} \to \frac{\bar{\lambda}_{25}}{\bar{\lambda}_{25}- 2\lambda_{25}} \, .  
\qquad ({\rm for} \ \nu_\chi \to 0)
\end{align}

To obtain $\kappa_\gamma \to 1$, in this limit we would require either that both $\kappa_3, \kappa_5 \to 0$, or for a cancellation to occur between the amplitude contributions from $H_3^+$ and $H_5^{+, ++}$.  Neither of these is possible in the $\nu_\chi \to 0$ limit.  Instead, the requirement of very small $\lambda_{25}$ needed to suppress the $h \to SS$ width in this limit drives both $\kappa_3$ and $\kappa_5$ to 1.  One way around this would be to take $\bar{\lambda}_{25}$ to be even smaller than $\lambda_{25}$, in which case $\kappa_5 \to 0$ while $\kappa_3 \to -2$, which still leads to a non-SM rate for $h \to \gamma\gamma$.  This latter possibility however drives $m_3$ and $m_5$ to zero and is hence precluded by the experimental lower bounds on $m_3$ and $m_5$ of 76~GeV from a combination of LEP-2 and ATLAS data (see Appendix~\ref{app:details} for details).

In the following section, we apply the Higgs total width and $h \to \gamma\gamma$ signal rate constraints quantitatively to show that the region with $m_S < m_h/2$ is entirely excluded at more than the $3\sigma$ level, thereby setting an experimental lower bound on $m_S$, and hence on $\nu_\chi$.  We will then proceed to study the constraints on the remaining parameter space of the Z2GM model.

\section{Experimental status of the Z2GM model}
\label{sec:expSTAT}

We now proceed to perform full numerical scans of the parameter space of the Z2GM model subject to the theoretical constraints from perturbative unitarity and vacuum stability.  We then sequentially apply the experimental constraints from direct searches for the non-SM Higgs bosons in the Z2GM model and LHC measurements of the signal strengths of the 125~GeV Higgs boson $h$.  Details of the scan procedure and the sources and implementation of the theoretical and experimental constraints are collected in Appendix~\ref{app:details}.

\subsection{Experimental status of the region with $m_S < m_h/2$}
\label{subsec:expLIGHT}

\begin{figure*}[b!]
   \resizebox{0.32\linewidth}{!}{\includegraphics{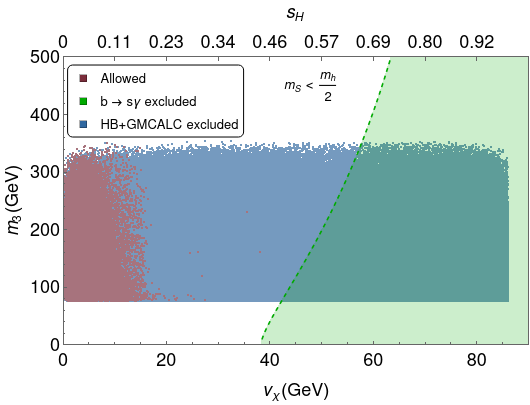}}
   \resizebox{0.32\linewidth}{!}{\includegraphics{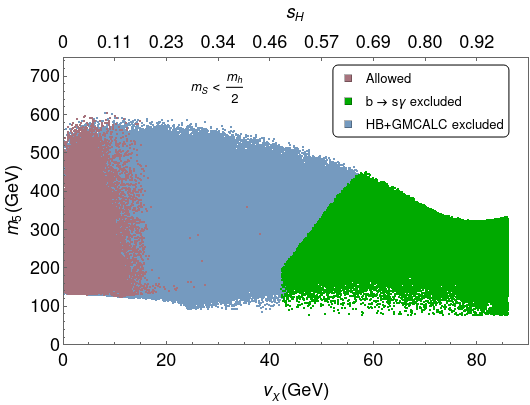}}
   \resizebox{0.32\linewidth}{!}{\includegraphics{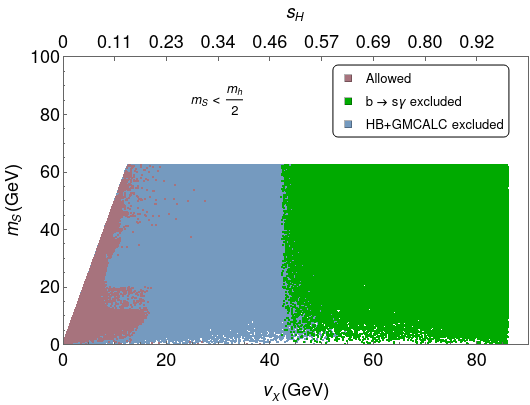}}
     \caption{ \label{fig:lightboundsM} Full parameter region with $m_S < m_h/2$ subject to theoretical constraints and the lower bounds $m_3, m_5 \geq 76$~GeV.  Green points are excluded by the constraint on $b \to s \gamma$, which eliminates the high-$\nu_\chi$ region in the $m_3$--$\nu_\chi$ plane (left plot).  Blue points are excluded by direct searches for the non-SM Higgs bosons in the Z2GM model, as implemented in GMCALC and HiggsBounds, together with the requirement that the 125~GeV Higgs total width is below the indirect bound of 19.1~MeV as discussed in the previous section. The maroon points survive the above constraints (but will be eliminated by the 125~GeV Higgs coupling measurements in Fig.~\ref{fig:exclusionlight}).}
\end{figure*}

We begin with a quantitative analysis of the parameter region with $m_S < m_h/2$.  To characterize the phenomenology of this region, we first show in Fig.~\ref{fig:lightboundsM} the full parameter region with $m_S < m_h/2$ subject to the theoretical constraints and the experimental lower bound $m_3, m_5 \geq 76$~GeV, which comes from a combination of LEP-2 searches for charged Higgs boson pair production (constraining $m_3$) and an ATLAS search at 8~TeV for like-sign dimuon production (constraining $m_5$).  

Applying the direct searches for non-SM Higgs bosons implemented in HiggsBounds and GMCALC, together with the indirect 99.7\% CL upper bound on the 125~GeV Higgs total width of 19.1~MeV discussed in the previous section, eliminates all but the maroon points in Fig.~\ref{fig:lightboundsM}.  Additionally, we can see that the $b\rightarrow s \gamma$ bound is redundant in this case, signalling the power of the direct searches in this region. The remaining parameter space is clustered at small $\nu_\chi$, mainly due to the suppression of the single-production cross sections of the additional Higgs bosons through vector boson fusion, gluon fusion, and fermion-antifermion fusion at small $\nu_\chi$, allowing them to evade the direct searches.  

While the masses $m_3$, $m_5$, and $m_S$ still individually populate most of their theoretically-allowed ranges, the bound on the 125~GeV Higgs total width strongly restricts the internal parameters of the model.  Because the contributions of the singly- and doubly-charged scalars to the $h \to \gamma\gamma$ rate involve different combinations of the same parameters that control $h \to SS$, imposing the experimental constraint on the $h \to \gamma\gamma$ rate will serve to entirely exclude this remaining region of the parameter space. 

\begin{figure}[b!]
 \resizebox{0.60\linewidth}{!}{\includegraphics{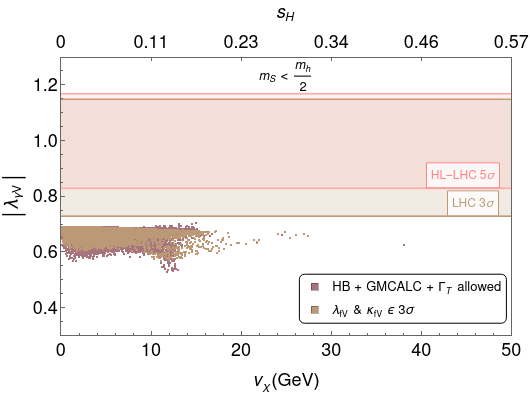}}
  \caption{\label{fig:exclusionlight} Surviving parameter space (maroon) from Fig.~\ref{fig:lightboundsM} as a function of the coupling ratio $|\lambda_{\gamma V}|$ and $\nu_\chi$.  The brown points show the surviving parameter space after imposing the $3\sigma$ constraints on $\lambda_{fV}$ and $\kappa_{fV}$.  The beige shaded stripe shows the $3\sigma$ allowed ATLAS range for this ratio; since no points remain in this range, the entire parameter region with $m_{S}<m_{h}/2$ is excluded.  The pink shaded stripe shows the $5\sigma$ projected sensitivity to this coupling ratio at the HL-LHC for comparison.}
\end{figure} 

To do this quantitatively, we use the ATLAS fit of ratios of Higgs coupling modifiers from Ref.~\cite{ATLAS2019}, as described in Appendix~\ref{app:details}.  Combining what amount to independent measurements of the same underlying parameters in the Z2GM model, we obtain bounds on the combinations $\lambda_{\gamma V} \equiv  \kappa_{\gamma} / \kappa_V$, $\lambda_{f V} \equiv \kappa_f / \kappa_V$, and $\kappa_{fV} \equiv \kappa_f \kappa_V / \kappa_H$, where $\kappa_H$ is defined in terms of the total Higgs width according to $\Gamma_T = \kappa_H^2 \Gamma_T^{SM}$.  

In order to visualize the exclusion, we plot the surviving points from Fig.~\ref{fig:lightboundsM} (in maroon) as a function of $|\lambda_{\gamma V}|$ and $\nu_\chi$ in Fig.~\ref{fig:exclusionlight}.  We further require that $\lambda_{fV}$ and $\kappa_{fV}$ lie within their $3\sigma$ (99.7\% CL) allowed ranges, shown in light brown; this restriction preserves most of the previously allowed parameter region in this projection.  However, we see that the allowed parameter space exhibits a very sizeable deviation from the SM in $\lambda_{\gamma V}$, with the allowed points lying in the range $|\lambda_{\gamma V}| \sim 0.5$--0.7.  This is mainly due to the suppression of $\kappa_\gamma$ caused by the additional singly- and doubly-charged Higgs bosons in the loop.  \emph{All} remaining points lie outside the current $3\sigma$ allowed range of $|\lambda_{\gamma V}|$, as shown by the beige shaded stripe in Fig.~\ref{fig:exclusionlight}.  For completeness, we also show as a pink shaded stripe the projected HL-LHC $5\sigma$ sensitivity to this coupling ratio~\cite{HLLHC}, which will exclude the entire parameter region with $m_S < m_h/2$ even more strongly if no deviation from the SM is found.

The exclusion of the entire parameter region with $m_{S} < m_{h}/2$ entirely excludes the possibility of very small $\nu_\chi$, setting a lower bound of $\nu_{\chi} > m_h/2\sqrt{8 \pi} \simeq 12.5$~GeV (or equivalently, $s_H \gtrsim 0.143$).
This experimental lower bound on $\nu_\chi$ in turn forces the couplings of $h$ to deviate non-negligibly from the SM.  This can be easily understood by fixing $\kappa_f = 1$ and trying to see how close to 1 one can make $\kappa_V$.  For $m_S < m_h$, $\kappa_f = 1$ requires $s_{\alpha} = \nu_\phi / \nu$.  Plugging this into the expression for $\kappa_V$ and simplifying yields
\begin{equation}
	\kappa_V = 1 - 8 \left( 1 \pm \sqrt{\frac{8}{3}} \right) \frac{\nu_\chi^2}{\nu^2} 
		\qquad ({\rm for} \ \kappa_f = 1),
\end{equation}
where the $\pm$ accounts for the two possible quadrants of the mixing angle $\alpha$.  This last expression is valid for both $m_S < m_h$ and $m_S > m_h$.  This means that it is impossible to achieve both $\kappa_f = 1$ and $\kappa_V = 1$ unless $\nu_\chi \to 0$, a possibility which we have just excluded.

We now turn to the analysis of the rest of the parameter space to further investigate the consequences of the lower bound on $\nu_\chi$.

\subsection{Experimental status of the rest of the parameter space}
\label{subsec:expREST}

\begin{figure*}[b!]
   \resizebox{0.32\linewidth}{!}{\includegraphics{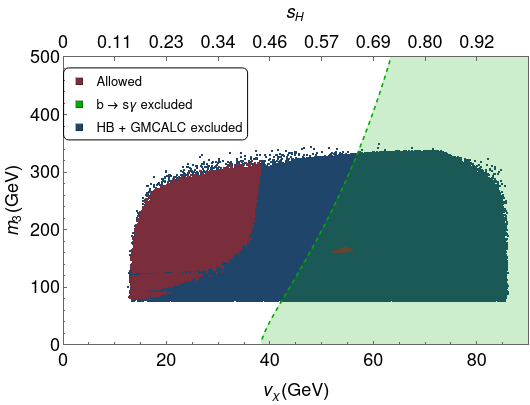}}
   \resizebox{0.32\linewidth}{!}{\includegraphics{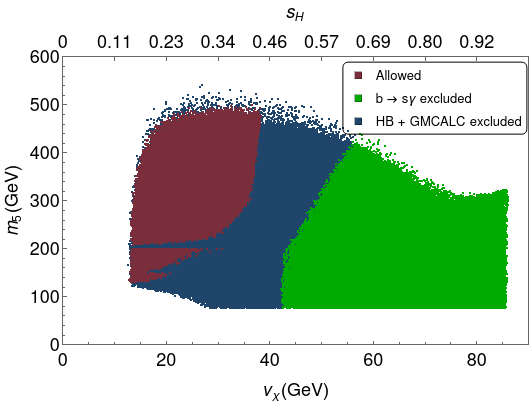}}
   \resizebox{0.32\linewidth}{!}{\includegraphics{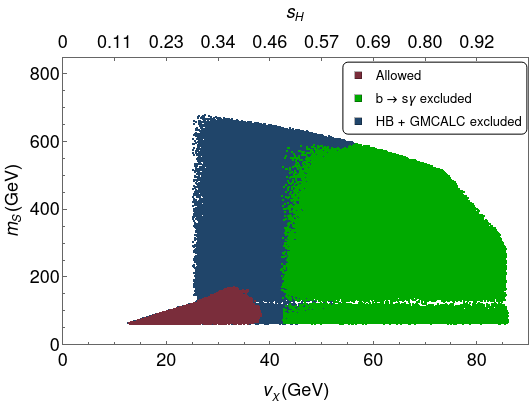}}
     \caption{ \label{fig:parameterfinal0} Full parameter region subject to theoretical constraints and the lower bounds $m_3, m_5 > 76$~GeV and $m_S > m_h/2$.  Green points are excluded by the constraint on $b \to s \gamma$, which eliminates the high-$\nu_\chi$ region in the $m_3$--$\nu_\chi$ plane (left plot).  Blue points are excluded by direct searches for the non-SM Higgs bosons in the Z2GM model, as implemented in GMCALC and HiggsBounds.  The maroon points survive these constraints.}
\end{figure*}

We begin by plotting in Fig.~\ref{fig:parameterfinal0} the masses $m_3$, $m_5$, and $m_S$ as a function of $\nu_\chi$ over the entire parameter space of the Z2GM model as allowed by theoretical constraints, subject to the experimental lower bounds $m_3, m_5 \geq 76$~GeV and the new bound found in the previous section $m_S > m_h/2$.  The lower bound on $\nu_\chi$ of about 12.5~GeV, imposed by the experimental constraint $m_S > m_h/2$, is clearly visible.  Points with large values of $\nu_\chi$ are excluded by $b \to s \gamma$; because the new-physics contribution to $b \to s \gamma$ depends only on $m_3$ and $\nu_\chi$, we show the excluded region shaded in green in the left panel of Fig.~\ref{fig:parameterfinal0}, while in the remaining panels the points excluded by $b \to s \gamma$ are shown in green.  The interplay between the $b \to s \gamma$ constraint and the upper bound on $m_3$ from perturbative unitarity and vacuum stability constraints in the Z2GM model entirely excludes $\nu_\chi$ values above about 58~GeV.

\begin{figure*}[b!]
   \resizebox{0.32\linewidth}{!}{\includegraphics{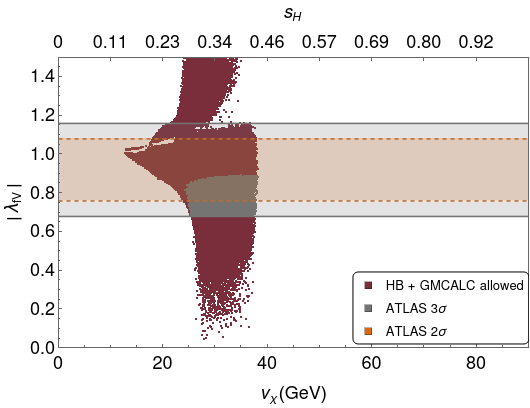}}
   \resizebox{0.32\linewidth}{!}{\includegraphics{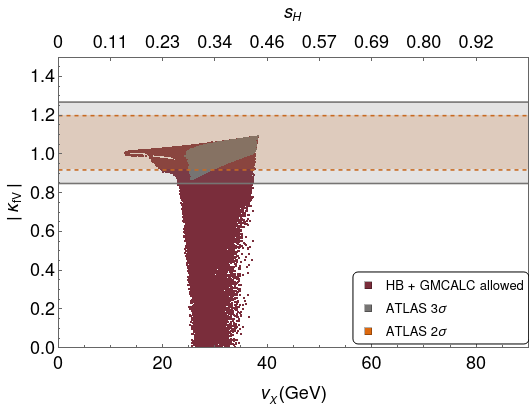}}
   \resizebox{0.32\linewidth}{!}{\includegraphics{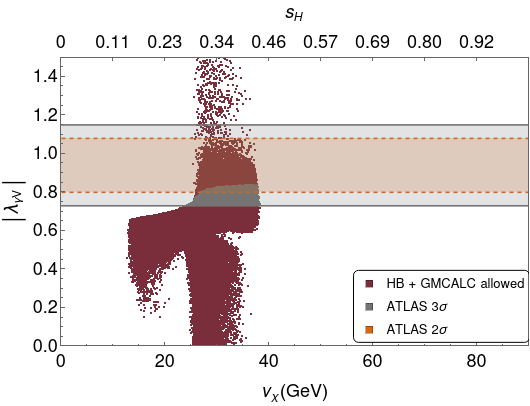}}
   \resizebox{0.32\linewidth}{!}{\includegraphics{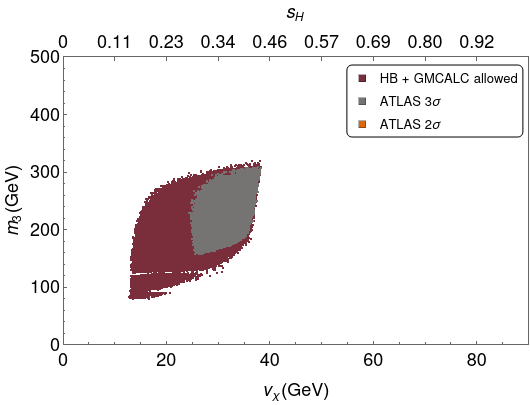}}
   \resizebox{0.32\linewidth}{!}{\includegraphics{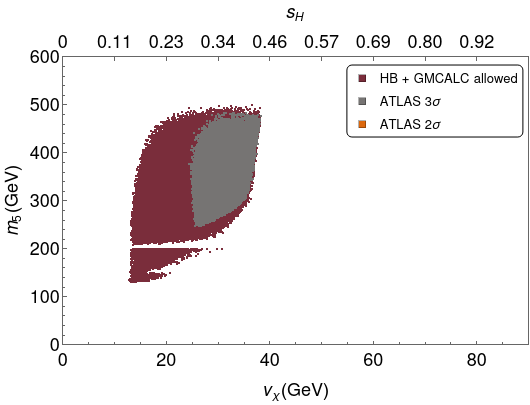}}
   \resizebox{0.32\linewidth}{!}{\includegraphics{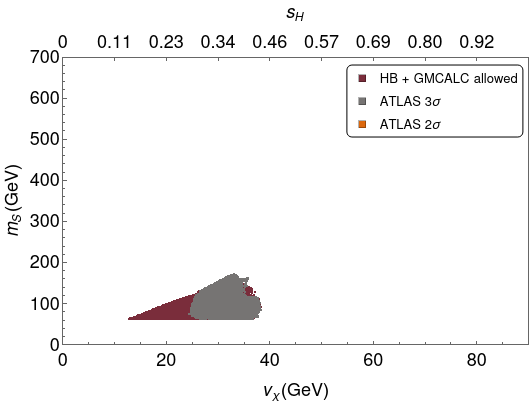}}
     \caption{ \label{fig:parameterfinal1} Surviving parameter space after applying direct searches for new Higgs bosons in the Z2GM model (the maroon points from Fig.~\ref{fig:parameterfinal0}), showing the ranges of the coupling modifier ratios $|\lambda_{fV}|$, $|\kappa_{fV}|$, and $|\lambda_{\gamma V}|$ (upper plots), as well as the masses $m_3$, $m_5$, and $m_S$ (lower plots), as a function of $\nu_\chi$.  The $2\sigma$ and $3\sigma$ allowed ranges of the coupling modifier ratios from ATLAS are shown as horizontal bands in the upper plots.  Points allowed at $3\sigma$ by all three coupling modifier ratio measurements are shown in grey.  No points are allowed at $2\sigma$ by all three coupling modifier ratios.}
\end{figure*}

We then apply the constraints from direct searches for non-SM Higgs bosons using the searches implemented in GMCALC and HiggsBounds.  The points excluded by these searches are shown in blue.  The most important of these searches in GMCALC is the search for vector boson fusion production of $H_5^{\pm\pm}$ for masses above 200~GeV with decays into like-sign $W$ boson pairs~\cite{CMS:2021wlt}, which together with the upper bound on $m_5$ from perturbative unitarity and vacuum stability constraints in the Z2GM model entirely excludes $\nu_\chi$ values above about 40~GeV; a search for Drell-Yan production of $H_5^{++}H_5^{--}$ with decays to like-sign $W$ pairs~\cite{ATLASnewIM}, which leads to an even stronger constraint for $m_5$ between 200 and 300~GeV; and theorist-recast constraints on $H_5^{\pm\pm} \to W^{\pm}W^{\pm}$~\cite{H5WW2} and $H_5^0 \to \gamma\gamma$~\cite{Ismail:2020zoz,Ismail:2020kqz}, which constrain the parameter space for $m_5$ below 200~GeV.  The sculpting of the surviving parameter space (shown in maroon) by these constraints on $H_5$ can be clearly seen in the middle panel of Fig.~\ref{fig:parameterfinal0}.  The most important of the searches implemented through HiggsBounds are LHC searches for $S$ decaying into $WW$, $ZZ$, or $hh$, which combine with the searches implemented in GMCALC to exclude $m_S$ values above about 175~GeV, and LEP searches for $e^+e^- \to SZ$ which are important when $m_S < m_h$.  Additional details of the direct searches are given in Appendix~\ref{app:details}.

We now consider the constraints from the 125~GeV Higgs boson coupling measurements.  Within the parameter space that survives the direct searches for additional Higgs bosons, very large deviations of the $h$ couplings from their SM values are possible.  These are shown in the upper three plots of Fig.~\ref{fig:parameterfinal1}, where we plot the absolute values of $\lambda_{fV} \equiv \kappa_f / \kappa_V$, $\kappa_{f V} \equiv \kappa_f \kappa_V / \kappa_H$, and $\lambda_{\gamma V} \equiv \kappa_{\gamma} / \kappa_V$ as a function of $\nu_\chi$.  The $2\sigma$ and $3\sigma$ allowed ranges of these observables based on the ATLAS Higgs coupling fit of Ref.~\cite{ATLAS2019} are shown respectively by the orange and grey horizontal stripes (see Appendix~\ref{app:details} for details of our choice and handling of this coupling fit).\footnote{A recent unpublished update~\cite{ATLAS2021} of the ATLAS analysis using more data gives a higher central value for $\lambda_{\gamma V}$ than the published results in Ref.~\cite{ATLAS2019}, yielding an even stronger exclusion of the model by about one additional standard deviation.}  Points that fall within the $3\sigma$ range of all three coupling combinations are shown in grey.  No points fall within the $2\sigma$ range of all three coupling combinations, which means that the Z2GM model is \emph{entirely} excluded at the $2\sigma$ level based on the Higgs coupling measurements of Ref.~\cite{ATLAS2019}.  The effect of the Higgs coupling constraints on the allowed ranges of $m_3$, $m_5$, and $m_S$ is shown in the lower three plots of Fig.~\ref{fig:parameterfinal1}.\footnote{Updated LHC analyses~\cite{ATLAS:2018gfm,CMS:2019bfg,CMS:2020osd} of charged Higgs production in top quark decays that have not yet been included in HiggsBounds were shown in Ref.~\cite{Ghosh:2022wbe} to exclude $\nu_\chi$ values as low as 10~GeV for $m_3$ between 90 and 130~GeV.  These exclusions would further constrain the low-$m_3$ region shown in maroon in the lower left panel of Fig.~\ref{fig:parameterfinal1}; this region is in any case also excluded by the 125~GeV Higgs coupling measurements.}

\begin{figure*}[b!]
   \resizebox{0.32\linewidth}{!}{\includegraphics{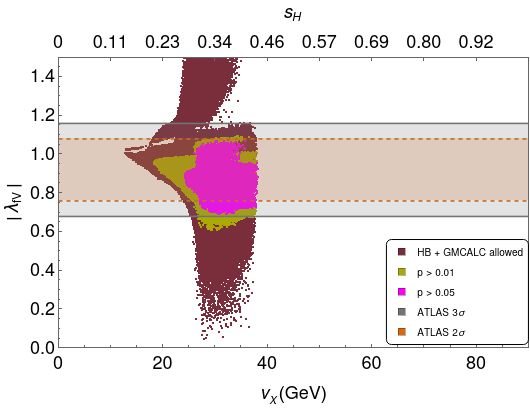}}
   \resizebox{0.32\linewidth}{!}{\includegraphics{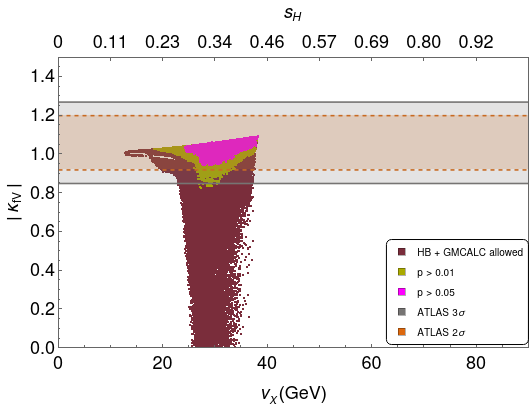}}
   \resizebox{0.32\linewidth}{!}{\includegraphics{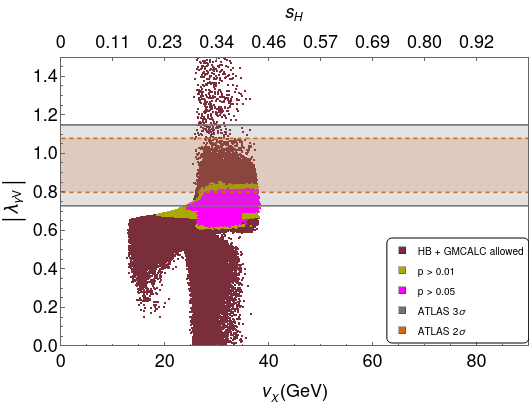}}
   \resizebox{0.32\linewidth}{!}{\includegraphics{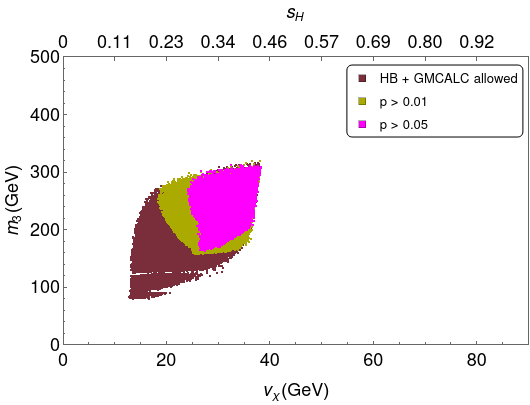}}
   \resizebox{0.32\linewidth}{!}{\includegraphics{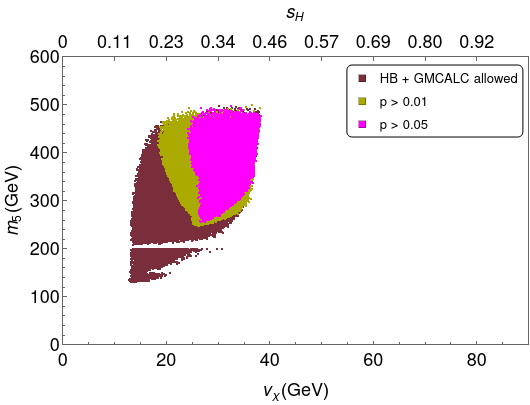}}
   \resizebox{0.32\linewidth}{!}{\includegraphics{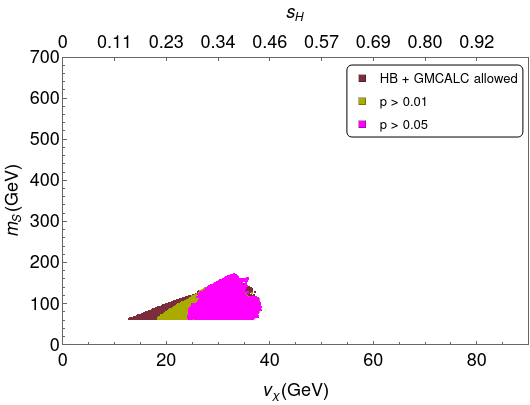}}
     \caption{ \label{fig:parameterfinal2} As in Fig.~\ref{fig:parameterfinal1} but showing the constraints from a HiggsSignals fit of $h$ signal strengths.  Magenta points have $p > 0.05$ (allowed at 95\% CL) and yellow points have $p > 0.01$ (allowed at 99\% CL).}
\end{figure*} 

As a cross-check, instead of directly applying the constraints from the coupling ratios we instead perform the fit of $h$ signal strengths to ATLAS and CMS data as implemented in HiggsSignals.  The advantage of using HiggsSignals is that it includes CMS data, as well as a more up-to-date collection of experimental inputs than the ATLAS coupling fit of Ref.~\cite{ATLAS2019}.  The disadvantages of using HiggsSignals are that the fit is less transparent, returning a $p$-value rather than providing insight into which observables are driving any discrepancies with experimental data, and that the statistical treatment implemented in HiggsSignals necessarily handles potentially correlated systematic uncertainties in a less sophisticated way than the dedicated coupling fits performed by the experiments themselves.  We show the results of the HiggsSignals fit in Fig.~\ref{fig:parameterfinal2} in the same form as Fig.~\ref{fig:parameterfinal1}, except that we show points with $p > 0.05$ in magenta and points with $p > 0.01$ in yellow.  HiggsSignals finds a small but not-insignificant parameter region with $p > 0.05$ (allowed at 95\% CL).  For this reason, we conclude that the Z2GM model is \emph{on the edge} of being excluded, but cannot yet be said to be \emph{fully} excluded.  Notice that the remaining allowed region from the HiggsSignals fit has $\lambda_{\gamma V} \sim 0.6$--0.85, which is considerably smaller than the SM value and will be further tested as the LHC collects additional data.

\begin{figure*}[t!]
   \resizebox{0.32\linewidth}{!}{\includegraphics{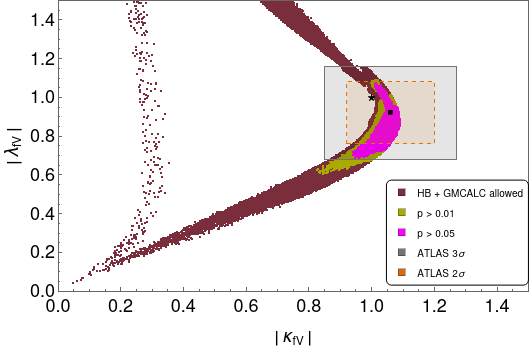}}
   \resizebox{0.32\linewidth}{!}{\includegraphics{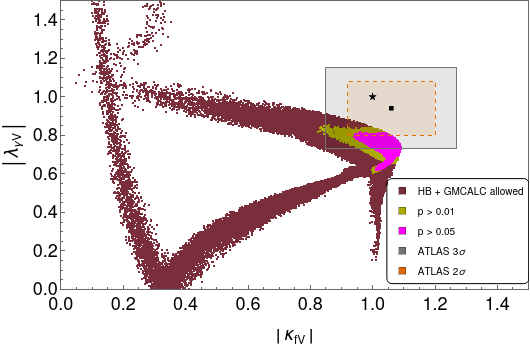}}
   \resizebox{0.32\linewidth}{!}{\includegraphics{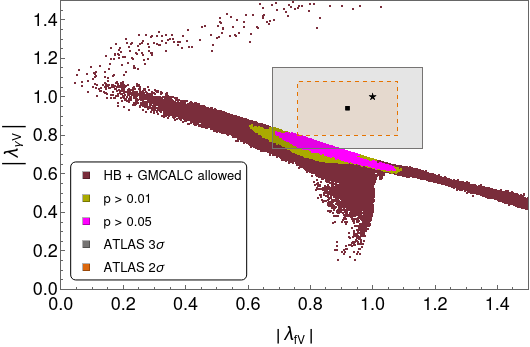}}
     \caption{ \label{fig:correlations} Correlations among the coupling modifier ratios.  Colours are the same as in Fig.~\ref{fig:parameterfinal2}.  The central values for the ATLAS coupling fit are indicated with a small black square and the SM prediction of (1,1) with a small black star.}
\end{figure*}

We can get a better understanding of the effect of the Higgs coupling fit by examining the correlations among the coupling modifier ratios.  To that end, in Fig.~\ref{fig:correlations} we plot pairs of the coupling modifier ratios $|\lambda_{fV}|$, $|\kappa_{fV}|$, and $|\lambda_{\gamma V}|$ against each other.  The colour scheme is the same as in Fig.~\ref{fig:parameterfinal2}.  We see that none of the pairs of coupling modifier ratios can be simultaneously SM-like (the black star at (1,1) in each panel of Fig.~\ref{fig:correlations}), though $|\lambda_{fV}|$ and $|\kappa_{fV}|$ come close.  $|\lambda_{\gamma V}|$, in particular, is rather far from being simultaneously SM-like with either of the other two coupling ratios; indeed, as shown in the rightmost panel of Fig.~\ref{fig:correlations}, none of the surviving parameter points are simultaneously within the $2\sigma$ allowed ranges of $|\lambda_{\gamma V}|$ and $|\lambda_{fV}|$.  This explains the absence of orange points in Fig.~\ref{fig:parameterfinal1}.  This behaviour ultimately derives from the non-decoupling nature of the Z2GM model.

\begin{figure}[h!]
 \resizebox{0.48\linewidth}{!}{\includegraphics{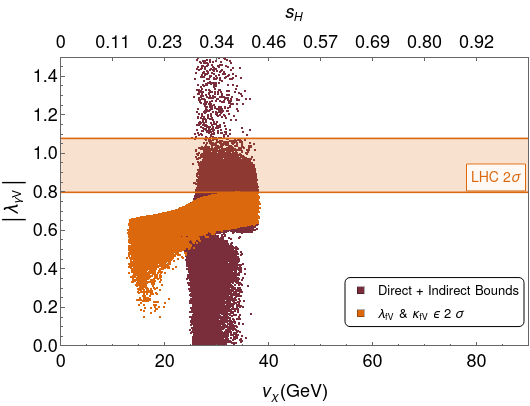}}  \hspace{0.02\linewidth} \resizebox{0.48\linewidth}{!}{\includegraphics{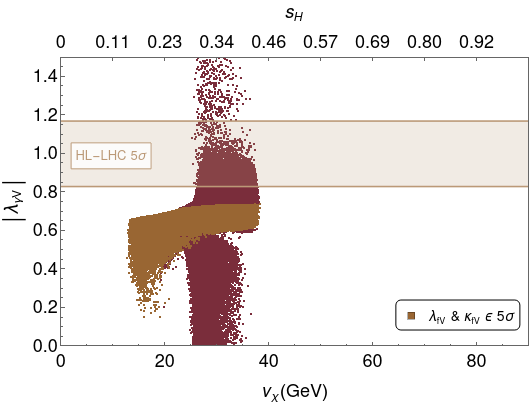}}
  \caption{\label{fig:HLPROJ} (Left) Surviving parameter space after applying direct searches for new Higgs bosons in the Z2GM model (the maroon points from Fig.~\ref{fig:parameterfinal0}), and in orange those points for which $\lambda_{fV}$ and $\kappa_{fV}$ are both within their $2\sigma$ experimental ranges.  The horizontal shaded orange band shows the $2\sigma$ allowed range of $|\lambda_{\gamma V}|$, showing that (just barely) no points satisfy all three coupling ratio constraints. 
(Right) The projection for the HL-LHC showing those points for which $\lambda_{fV}$ and $\kappa_{fV}$ are both within their expected $5\sigma$ ranges, along with the projected $5\sigma$ allowed range of $|\lambda_{\gamma V}|$, assuming that no deviations from the SM are found, showing that the HL-LHC will decisively exclude the entire Z2GM model in this scenario.}
\end{figure}

We finally consider the prospects for improved measurements of the Higgs coupling modifier ratios at the HL-LHC~\cite{HLPROJATLAS}.  In the left panel of Fig.~\ref{fig:HLPROJ} we reproduce the magenta points from the upper right panel of Fig.~\ref{fig:parameterfinal1}, showing also in orange the points for which $\lambda_{fV}$ and $\kappa_{fV}$ lie within their $2\sigma$ ranges from the ATLAS coupling fit~\cite{ATLAS2019}.  The horizontal orange stripe shows the $2\sigma$ allowed range for $|\lambda_{\gamma V}|$, which is plotted on the $y$-axis.  As shown before, the model is just barely excluded at $2\sigma$ by this analysis.  For comparison, in the right panel of Fig.~\ref{fig:HLPROJ} we plot the points that survive the current LHC constraints from direct searches and are within the projected $5\sigma$ range of $\lambda_{fV}$ and $\kappa_{fV}$ at the HL-LHC~\cite{HLPROJATLAS}, assuming that their experimental central values will be SM-like.  The horizontal shaded brown band in this plot shows the projected $5\sigma$ range for $|\lambda_{\gamma V}|$.  The fact that the points are well outside this band indicates that the HL-LHC will decisively exclude the entire Z2GM model assuming that no deviations of these couplings from their SM values are found.  In this case, if the GM model occurs in nature, an explicit breaking of the $Z_2$ symmetry would be required.
 
\section{Conclusions} \label{sec:conc}

In this paper, we analyzed the phenomenology of a constrained version of the Georgi-Machacek (GM) model with a $Z_2$ symmetry enforced in the scalar potential (Z2GM).  Unlike the full GM model, when the triplet vacuum expectation value $\nu_{\chi}$ is nonzero the Z2GM model does \emph{not} possess a decoupling limit in which the additional scalars can all be taken heavy while the couplings of the remaining 125~GeV Higgs boson approach their SM values.  This implies that the Z2GM model will exhibit some irreducible deviations from the SM, and could potentially be entirely excluded by current or near-future experiments.  

A key component of this analysis stemmed from the fact that in the limit of small $\nu_\chi$ in the Z2GM model, the second custodial singlet $S$ becomes very light, opening the possibility of a significant decay width of $h \to SS$.  The limited parameter freedom of the Z2GM model also prevents a simultaneous suppression of the $hSS$ coupling and the $h$ couplings to the singly- and doubly-charged scalars which modify the loop-induced $h \gamma\gamma$ coupling.  These two features allow current constraints on the Higgs total width and the $h \to \gamma\gamma$ rate to entirely exclude the parameter region with $m_S < m_h/2$ at $99.7\%$ confidence level.  This in turn puts an absolute lower bound on $\nu_\chi$ of about 12.5~GeV.

This lower bound on $\nu_\chi$ \emph{enforces} a nonzero minimal deviation of the tree-level couplings of $h$ from their SM values.  The loop-induced coupling of $h$ to $\gamma\gamma$ also receives significant modifications from the presence of the non-decoupling singly- and doubly-charged scalars in the loop.  Combining constraints on the model from direct searches for non-SM Higgs bosons and measurements of the couplings of the 125~GeV Higgs boson, we showed that the Z2GM model is on the verge of being fully excluded by current experimental data.  The remaining parameter space exhibits sizeable deviations in the 125~GeV Higgs couplings, particularly in $\lambda_{\gamma V} \equiv \kappa_\gamma / \kappa_V$, which is suppressed by 15--40\% compared to its SM value.  It also restricts $S$ to be lighter than about 175~GeV, which could be probed by a future $e^+e^-$ collider.  We show that the addition of the projected HL-LHC measurements of the 125~GeV Higgs couplings will completely exclude the model at more than the $5\sigma$ level, assuming that no deviation from the SM is found.  In this case, if the GM model occurs in nature, an explicit breaking of the $Z_2$ symmetry would be required.

\acknowledgements
We thank Jo\~ao G. Alencar Carib\'e for helpful comments and discussions. This work was supported by the Natural Sciences and Engineering Research Council of Canada (NSERC).

\appendix

\section{\label{sec:appendix} Generalization of the mass hierarchy relation} 

The relation derived in the Z2GM model for the mass hierarchy between $h$ and $S$, Eq.~\eqref{eq:hierachy}, can be generalized for an arbitrary $2 \times 2$ hermitian matrix of which one eigenvalue is fixed. This means that we can also do this analysis for the GM model without the extra $Z_2$ symmetry and compare the results to the more constrained Z2GM case.  In this appendix, we derive this result generically and then apply it to both the Z2GM and GM models to illustrate the unique feature created by the $Z_{2}$ symmetry. The core of the argument is the level repulsion phenomenon for matrix eigenvalues. 

Let $\Lambda$ be a $2 \times 2$ hermitian matrix that has the general form:
\begin{align}
\Lambda = \begin{pmatrix}
\lambda_{11} & \lambda_{12} \\
\lambda_{12}^{*} & \lambda_{22}
\end{pmatrix} \, .
\end{align}
The eigenvalues of the matrix will be called $\lambda_{a}$ and $\lambda_{b}$. In principle, they can be any real numbers. However, for our setup, we want to fix $\lambda_{a}$ to be a specific value, $\lambda_{fix}$. We can implement this by adjusting $\lambda_{11}$ such that $\lambda_{a}=\lambda_{fix}$. This procedure is uniquely determined and does not depend on which eigenvalue is larger or smaller:
\begin{align}
\lambda_{11} = \lambda_{fix} + \frac{\abs{\lambda_{12}}^{2}}{\lambda_{22}-\lambda_{fix}} \, .
\end{align}

We then insert this relation into the trace of $\Lambda$:
\begin{align}
\Tr(\Lambda) &= \lambda_{11}+\lambda_{22}  
=\lambda_{fix} + \frac{\abs{\lambda_{12}}^{2}}{\lambda_{22}-\lambda_{fix}} + \lambda_{22}
= \lambda_{fix} + \lambda_{b}.
\end{align}

This gives us a unique result for the other eigenvalue $\lambda_{b}$:
\begin{align}
\lambda_{b} = \lambda_{22} + \frac{\abs{\lambda_{12}}^{2}}{\lambda_{22}-\lambda_{fix}}   \, .
\end{align}
Given this relation we can subtract $\lambda_{fix}$ from both sides to obtain the hierarchy between the eigenvalues:
\begin{align} \label{eqeq}
\lambda_{b} - \lambda_{fix} = \mathcal{K} + \frac{\abs{\lambda_{12}}^{2}}{\mathcal{K}} \, , \qquad
{\rm where} \ \mathcal{K} = \lambda_{22} - \lambda_{fix} \, .
\end{align}

This result is the generalization of what we obtained in Eq.~\eqref{eq:hierachy}. Since the matrix $\Lambda$ is hermitian, the numerator of the second term in Eq.~\eqref{eqeq} is always positive. This means that the sign of the right-hand side is controlled by the sign of $\mathcal{K}$. The eigenvalue $\lambda_b$ is larger than the fixed one $\lambda_{fix}$ when $\mathcal{K}>0$ and smaller when $\mathcal{K}<0$. 

We can understand why this result happens if we look at the matrix when $\lambda_{12}=0$. In this case the eigenvalues are directly determined $\lambda_{11}=\lambda_{a}$, $\lambda_{22}=\lambda_{b}$. Now we adjust $\lambda_{11}$ such that $\lambda_{a} = \lambda_{fix}$. This give us the system with the eigenvalues $\lambda_{fix}$ and $\lambda_{22}=\lambda_{b}$. Now, turning on the off-diagonal term these eigenvalues will repel by an amount $\Delta$. This means that the new eigenvalues are $\lambda_{fix} + \Delta$ and $\lambda_{22}-\Delta$.  In this example we chose $\lambda_{fix}$ to be the larger eigenvalue; in the opposite situation, the sign of $\Delta$ will be flipped.

Because we want one of the eigenvalues to be equal to $\lambda_{fix}$, we again need to adjust $\lambda_{11}$ to enforce this. This changes the value of $\Delta$, but the second eigenvalue remains repelled from its initial value. This means that the hierarchy between the two eigenvalues is preserved when $\lambda_{12}$ is nonzero, simply because $|\lambda_{12}|^{2}$ is always positive.

We now use this result to compare the $m_{S}<m_{h}$ region in the Z2GM model to that in the GM model without the $Z_2$ symmetry.  Using the explicit expressions for the custodial-singlet mass matrix in the two models, we obtain 
\begin{align}
\mathcal{K}_{Z2GM} &= 8 \lambda_{34}\nu_{\chi}^{2}- m_{h}^{2} \, , \\
\mathcal{K}_{GM} &= \frac{M_{1} \nu_{\phi}^{2}}{4\nu_{\chi}} - 6 M_{2} \nu_{\chi} + 8 \lambda_{34}\nu_{\chi}^{2}- m_{h}^{2}  \, , \label{eq:KGM}
\end{align}
where the scalar potential for the GM model is identical to that in Eq.~\eqref{eq:pot} with the addition of the two $Z_2$-breaking terms,~\cite{litGM1}
\begin{equation}
	V_{GM} = V_{Z2GM} - M_1 {\rm Tr}(\Phi^{\dagger} \tau^a \Phi \tau^b) (U X U^{\dagger})_{ab}
	- M_2 {\rm Tr}(X^{\dagger} t^a X t^b) (U X U^{\dagger})_{ab}.
\end{equation}
Here $U$ is a unitarity matrix given in Ref.~\cite{litGM1} that rotates $X$ into the Cartesian basis.

\begin{figure}[h!]
 \resizebox{0.43\linewidth}{!}{\includegraphics{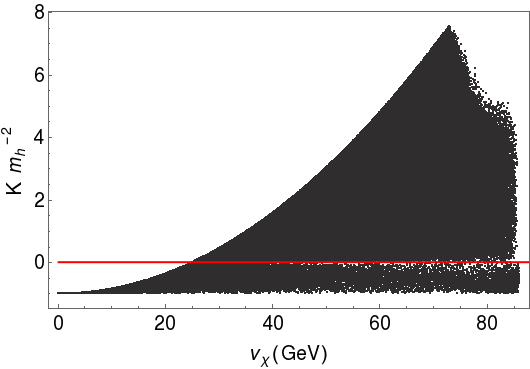}}
 \, \,
 \resizebox{0.45\linewidth}{!}{\includegraphics{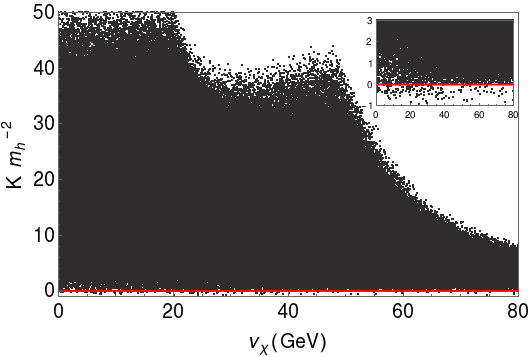}}
     \caption{\label{fig:k} Parameter space scans showing $\mathcal{K}$ as a function of $\nu_\chi$ for the Z2GM (left) and GM (right) models.  The red line indicates $\mathcal{K} = 0$, which divides the two mass hierarchies $m_S > m_h$ (positive $\mathcal{K}$) from $m_S < m_h$ (negative $\mathcal{K}$).}
\end{figure}

We plot $\mathcal{K}$ (normalized by $m_h^2$ to make it dimensionless) against $\nu_\chi$ in Fig.~\ref{fig:k} for the two models from scans over the model parameters.  The red line shows $\mathcal{K} = 0$ in order to highlight the region of negative $\mathcal{K}$ values, which give rise to the hierarchy $m_S < m_h$.  In the Z2GM model (left panel of Fig.~\ref{fig:k}), the region with small $\nu_\chi$ is populated \emph{only} with values of $\mathcal{K}$ less than zero.  This is enforced by the upper bound $\lambda_{34} < \pi$ from perturbative unitarity, which gives rise to the parabolic shape of the upper left edge of the populated parameter region.  In contrast, the GM model (right panel of Fig.~\ref{fig:k}) is well populated with positive values of $\mathcal{K}$ all the way down to $\nu_\chi = 0$ (negative values of $\mathcal{K}$ also appear, as shown in the inset).  This happens because the first term in Eq.~\eqref{eq:KGM} can easily be larger than $m_h^2$ even when $\nu_\chi$ is very small.

\section{Theoretical and experimental constraints applied}
\label{app:details}

In this appendix, we collect the details of the theoretical and experimental constraints applied to the Z2GM model in our analysis.  We also describe the strategy that we used to scan the parameter space.  In our analysis, we used the public codes GMCALC version 1.5.3~\cite{gmcalc}, HiggsBounds version 5.3.2 beta~\cite{HiggsBounds}, and HiggsSignals version 2.2.3~\cite{HiggsSignals}.  
HiggsBounds and HiggsSignals were called from within GMCALC.

\subsection{Theoretical and indirect constraints}

We require that the parameters of the scalar potential satisfy the constraints from perturbative unitarity of two-to-two scalar scattering amplitudes~\cite{unitarity} and that the scalar potential is bounded from below~\cite{litGM1}.  While these conditions were originally found for the unconstrained GM model (without the additional $Z_2$ symmetry), they do not depend on the $Z_2$-breaking terms and hence also apply to the Z2GM model.  Together with the requirement that all the squared masses of physical scalars are positive, these conditions restrict the allowed ranges of the quartic couplings so that all allowed values are captured by the scan ranges for $\lambda_2$, $\lambda_3$, $\lambda_4$, and $\lambda_5$ described in Ref.~\cite{gmcalc}, with the additional constraint $\lambda_5 > 0$ in the Z2GM model from the requirement that $m_3^2 > 0$.  The full set of perturbative unitarity and boundedness-from-below constraints are then applied to each point by GMCALC.

We also apply a check in GMCALC, implemented using a numerical scan of vevs, that rejects points for which the potential possesses a deeper minimum in which the custodial symmetry is spontaneously broken~\cite{litGM1,Moultaka:2020dmb} (i.e., we require that the custodial-symmetric vacuum is the global minimum of the scalar potential).

The Z2GM model is also subject to constraints from $B$ physics, arising from one-loop processes involving the top quark and the charged scalar $H_3^+$.  These constraints depend only on $m_3$ and $\nu_\chi$ (or alternatively $s_H$), and the most important of them is from $b \to s \gamma$~\cite{Hartling:2014aga}, which is also applied via its implementation in GMCALC (we apply the ``loose'' bound as described in Ref.~\cite{Hartling:2014aga}).  This excludes parameter points with large values of $\nu_\chi$.  The same points will also be excluded by the direct searches for additional Higgs bosons that we discuss below.

Finally, in the analysis of the parameter region with $m_S < m_h/2$, we use the indirect bound on the 125~GeV Higgs boson total width from analyses of on- and off-shell production in the four-lepton final state~\cite{CMS2019},
\begin{align}
	\Gamma_{T} < 19.1~{\rm MeV}  \qquad {\rm at \ 99.7\% \ CL}.
\end{align}
This can be compared to the SM prediction for the 125~GeV Higgs boson total width of 4.09~MeV~\cite{LHCHiggsCrossSectionWorkingGroup:2016ypw}.  Such a large enhancement of the Higgs total width due to the non-SM decay $h \to SS$ would dramatically modify the Higgs branching ratios to SM final states and hence would also be excluded by Higgs signal strength measurements.

\subsection{Constraints from direct searches for additional Higgs bosons}

\subsubsection{Direct searches implemented in GMCALC}

GMCALC implements several direct searches for the additional Higgs bosons of the Z2GM model, including dedicated LHC searches for the doubly-charged Higgs $H_5^{\pm\pm}$ and searches recast by theorists to constrain the model.  These are as follows:

\begin{itemize}
\item Production of $H_5^{\pm\pm}$ in vector boson fusion (VBF) with decays to $W^{\pm} W^{\pm} \to$ like-sign dileptons.  The cross-section is proportional to $\nu_{\chi}^2$, so this process directly constrains large triplet vevs.  We include the latest dedicated CMS search for this process~\cite{CMS:2021wlt} valid for $m_5 \geq 200$~GeV, as well as a theorist recast~\cite{H5WW2} of an ATLAS search for like-sign $W$ boson production in VBF at 8~TeV~\cite{ATLAS:2014jzl} which constrains this process for $m_5$ as low as 100~GeV.  The implementation of these searches in GMCALC accounts for the fact that BR($H_5^{++} \to W^+ W^+$) can be less than one (due to $H_5 \to H_3 W$ and $H_5 \to H_3 H_3$ decays).

\item Drell-Yan production of $H_5^{++}H_5^{--}$ or $H_5^{\pm\pm}H_5^{\mp}$ with $H_5^{\pm\pm}$ decaying into like-sign $W$ boson pairs.  We include a recent dedicated ATLAS search~\cite{ATLASnewIM} valid for $m_5 \geq 200$~GeV that \emph{entirely} excludes $m_5$ values between 200 and 350~GeV under the assumption that BR($H_5^{++} \to W^+ W^+) = 1$; the implementation of this search in GMCALC accounts for the possibility that this branching ratio is less than one, which indeed occurs in the Z2GM model, so that some parameter space in this mass range survives this direct constraint.  We also include a theory recast~\cite{DYELLH52,DYELLH53} of ATLAS like-sign dimuon data at 8~TeV~\cite{DYELLH51}, which puts a lower bound on $m_5$ of 76~GeV under the assumption that BR($H_5^{++} \to W^+ W^+) = 1$; i.e., that $H_5^{\pm\pm} \to W^{\pm}H_3^{\pm}$ does not compete with decays of $H_5^{++}$ into like-sign $W$ pairs.  Together with LEP-2 searches for pair production of singly-charged Higgs bosons~\cite{LEPHiggsWorkingGroupforHiggsbosonsearches:2001ogs} (interpreted here as $e^+e^- \to H_3^+H_3^-$), which exclude the possibility of $m_3 < 78$~GeV assuming that $H_3^{\pm}$ decays entirely into a combination of $\tau \nu$ and $cs$ final states and hence remove the possibility of $H_5^{++}$ decays to $W^+ H_3^+$, this allows us to impose an \emph{absolute} lower bound $m_3, m_5 \geq 76$~GeV.

\item Drell-Yan production of $H_{5}^{0}H_{5}^{\pm}$ with $H_{5}^{0} \rightarrow \gamma \gamma$.  We include a theory recast~\cite{Ismail:2020zoz,Ismail:2020kqz} of an ATLAS diphoton resonance search at 8~TeV~\cite{DYELLH01}, which significantly constrains the model for $m_5$ below about 120~GeV.

\end{itemize}

\subsubsection{Direct searches from HiggsBounds}

HiggsBounds implements a very large number of direct search limits for neutral and singly-charged Higgs bosons from LEP, Tevatron, and LHC experiments.  We apply the HiggsBounds constraints to the additional Higgs bosons $S$, $H_3^0$, $H_5^0$, and $H_3^{\pm}$.  We do not apply the HiggsBounds constraints to the 125~GeV Higgs boson $h$ decays into SM final states because HiggsBounds bases its exclusions on applying the single most sensitive experimental analysis to any given model point; therefore, a downward fluctuation in the 125~GeV Higgs boson event rate in a single measurement could exclude model points that would more properly be allowed based on a global combination of Higgs signal strengths.  We will later use HiggsSignals to constrain the 125~GeV Higgs boson's production and decay rates in SM channels.

After applying the direct searches implemented in GMCALC as discussed above, the additional parameter regions excluded by HiggsBounds are almost entirely due to searches for the second custodial singlet $S$.  These include constraints from LEP searches for $e^{+} e^{-}\rightarrow S Z$ (with $S \to b\bar b$, $\gamma\gamma$, and inclusive final states) when $S$ is sufficiently light, as well as LHC searches involving $S \to WW$, $ZZ$, and $hh$ at heavier $S$ masses.

\subsection{Constraints from production and decay rates of the 125~GeV Higgs boson}

In the Z2GM model the couplings of $h$ to fermion pairs, $W$ and $Z$ boson pairs, and photon pairs are modified compared to their values in the SM.  We can therefore use LHC measurements of Higgs production and decay rates, which are sensitive to these couplings, to constrain the parameter space.  We use two different strategies to apply these constraints and compare their results in the text.

\subsubsection{Ratios of coupling modifiers}

In an ideal world, we would test each model point by applying the experimental constraints on the Higgs coupling modification factors $\kappa_f$, $\kappa_V$, and $\kappa_\gamma$.  This $\kappa$-framework is useful when we have new states around the electroweak scale, which is the case in the Z2GM model.  Unfortunately, direct fits of the Higgs boson couplings based on LHC data necessarily require assumptions to be made in order to eliminate flat directions.  In particular, the most common fits are made assuming modifications to $\kappa_f$ and $\kappa_V$ (but no new particles in the loops contributing to $\kappa_g$ or $\kappa_{\gamma}$), or modifications to $\kappa_g$ and $\kappa_\gamma$ due to new particles in the loops (but no modification of $\kappa_f$ and $\kappa_V$).  Because the Z2GM model predicts modifications to $\kappa_f$ and $\kappa_V$ along with new particles in the loops for $\kappa_\gamma$, we are restricted to using coupling fits that accommodate this possibility.  The only such coupling fit that exists is for the generic parameterization of six ratios of coupling modifiers $\lambda_{ij} \equiv \kappa_i/\kappa_j$ together with one overall measure of the signal rate $\kappa_{gZ} \equiv \kappa_g \kappa_Z/\kappa_H$, where $\kappa_H$ parameterizes modifications to the total width of the Higgs according to $\Gamma_T = \kappa_H^2 \Gamma_T^{SM}$.  We therefore take as input the most recent published fit by ATLAS in Table~12 of Ref.~\cite{ATLAS:2019nkf}.\footnote{An updated version of the same fit using additional data appeared recently in Table 9 of Ref.~\cite{ATLAS2021}, in which $\lambda_{\gamma V}$ fluctuates to a higher central value, making the model even more excluded based on the analysis of coupling-modifier ratios.}

For added statistical power we can take advantage of the fact that the Z2GM model obeys $\kappa_{t}=\kappa_{b}=\kappa_{\tau}=\kappa_{g} \equiv \kappa_{f}$ and $\kappa_{Z}=\kappa_{W} \equiv \kappa_{V}$, so that several of these coupling modifier ratios represent independent measurements of the same underlying combination of model parameters.  We statistically combine these ``redundant'' measurements assuming that the uncertainties are Gaussian distributed (we symmetrize asymmetric uncertainty ranges by taking their average) and ignoring the fact that some of the systematic uncertainties are correlated; in particular, we combine $\lambda_{\tau Z}$, $\lambda_{b Z}$ and $\lambda_{Zg}^{-1}$ to obtain $\lambda_{fV}$.  By this method we obtain,
\begin{align}
 \kappa_{fV} &= \frac{\kappa_{f}\kappa_{V}}{\kappa_{H}} = 1.06 \pm 0.07 \, , \\
  \lambda_{\gamma V} &= \frac{\kappa_{\gamma}}{\kappa_{V}} = 0.94 \pm 0.07 \, , \\
   \lambda_{fV} &= \frac{\kappa_{f}}{\kappa_{V}} = 0.92 \pm 0.08 \, .
 \end{align}
In our analysis of the Z2GM parameter space we require that each of these observables \emph{separately} lies within $2$ or $3\sigma$ of its central value; i.e., we do not combine their likelihoods.
 
We also consider the anticipated precision of these measurements at the HL-LHC~\cite{HLPROJATLAS}.  Combining channels, in the same way, we obtain the anticipated $1\sigma$ experimental uncertainties,  
\begin{align}
 \delta \kappa_{fV} = 0.034  \, , \qquad
  \delta \lambda_{\gamma V} = 0.024 \, , \qquad
   \delta \lambda_{fV} = 0.034 \, . \qquad {\rm (HL-LHC)}
 \end{align}

\subsubsection{HiggsSignals}

As an independent cross-check using experimental inputs from both ATLAS and CMS and different statistical methods, we also apply the global Higgs signal strength fit in HiggsSignals~\cite{HiggsSignals}.  We treat each parameter point as its own model (with zero free parameters) to extract the $p$-value.  We apply the peak-centred $\chi^{2}$ method implemented in HiggsSignals incorporating all neutral scalars in order to capture total signal rates when one of the other scalars is close in mass to the 125~GeV Higgs.  We show regions with $p > 0.05$ (allowed at 95\% CL) and $p > 0.01$ (allowed at 99\% CL).  Overall, the region of parameter space allowed by the HiggsSignals global fit is similar to that obtained using the coupling modifier fit based on ATLAS data alone, which gives us higher confidence in the robustness of our conclusions.

\subsection{Numerical scan procedure}

The Z2GM model contains 7 parameters; after fixing $G_F$ and $m_h$, we are left with 5 free parameters that must be scanned.  We choose the parameters $\mu_3^2$, $\lambda_2$, $\lambda_3$, $\lambda_4$, and $\lambda_5$ as the free parameters and fix $\mu_2^2$ and $\lambda_1$ in terms of these and the measured values of $G_F$ and $m_h$.  This corresponds to {\tt INPUTSET = 2} in GMCALC (we also set $M_1 = M_2 = 0$ in GMCALC in order to implement the $Z_2$ symmetry).  We also require that the global minimum has nonzero $\nu_{\chi}$; i.e., that the $Z_2$ symmetry is spontaneously broken.\footnote{In this paper we do not analyze the dark matter phase of the theory in which $\nu_\chi = 0$ and the $Z_2$ symmetry is preserved.}
All of our numerical scans begin by imposing the theoretical constraints (perturbative unitarity, boundedness from below, and the absence of deeper minima) together with the lower bound $m_3, m_5 \geq 76$~GeV from direct searches.

The scans are performed in two steps.  First, we use GMCALC to generate random samples uniformly distributed in the variables $\sqrt{|\mu_3^2|}$, $\lambda_2$, $\lambda_3$, $\lambda_4$, and $\lambda_5$ and apply the relevant constraints to discard excluded points.  Then, to better populate the allowed parameter regions and improve the efficiency of the scans, we use the surviving points as input to the machine learning implementation {\tt LearnDistribution} in MATHEMATICA~\cite{Mathematica}.  This allows us to efficiently generate a very large number of points in MATHEMATICA concentrated in the vicinity of the relatively small surviving regions of parameter space, which we finally feed back through GMCALC to calculate physical observables and impose the relevant constraints.

\bibliographystyle{apsrev-title}
\bibliography{bibGM2}

\end{document}